\documentclass{aastex631}

\usepackage{graphicx}
\usepackage{booktabs}
\usepackage{tabularx}
\newcolumntype{Y}{>{\centering\arraybackslash}X} 
\newcolumntype{Z}{>{\sisetup{table-format=1.2}}S}

\begin{document}

\title{From Empirical to Physical Model: Direct Fits of Optically Thin Inverse Compton Scattering to Prompt GRB Spectra}

\author{Pragyan Pratim Bordoloi}
\correspondingauthor{Pragyan Pratim Bordoloi}
\email{pragyan.bordoloi21@iisertvm.ac.in}
\affiliation{School of Physics, Indian Institute of Science Education and Research Thiruvananthapuram, Kerala, 695551, India\\}

\author{Shubh Mittal}
\affiliation{School of Physics, Indian Institute of Science Education and Research Thiruvananthapuram, Kerala, 695551, India\\}
\affiliation{Inter-University Centre for Astronomy and Astrophysics (IUCAA), Pune, Maharashtra,  411007, India\\}

\author{Shabnam Iyyani}%
\affiliation{School of Physics, Indian Institute of Science Education and Research Thiruvananthapuram, Kerala, 695551, India\\}
 \affiliation{%
 Centre of High Performance Computing, Indian Institute of Science Education and Research Thiruvananthapuram, Kerala, 695551, India\\
}



\begin{abstract}
Gamma-ray burst (GRB) prompt emission is commonly attributed to non-thermal radiation processes operating in the optically thin 
regions of a relativistic outflow. Among these, optically thin inverse-Compton (IC) scattering remains an important yet under-tested mechanism. From an initial set of 41 bursts selected 
using empirical Band-function criteria that highlight quasi-thermal low-energy slopes ($\alpha > -0.5$) and constrained 
high-energy indices ($-1.7 >\beta > -3.3$), only four events satisfy these conditions consistently in both time-integrated and time-resolved spectra. The IC fits yield self-consistent constraints on the seed-photon field and the electron population at the dissipation site. For bulk Lorentz factors $\Gamma \sim 170$ - 
$550$, we infer seed thermal peaks of $\sim 0.05$ - $0.2$ keV and electron thermal energies of $\sim 20$–$300$ keV in the co-moving frame. A fraction of only $0.1\%$–$20\%$ of electrons are accelerated into a non-thermal tail with an average index value of $\delta \sim 1.8$.  The derived Comptonisation parameters indicate moderate $y$ 
values ($\sim1$–$3$), optical depths $\tau \sim 0.2$–$0.6$, and dissipation radii just above the photosphere, consistent with 
mildly relativistic ($\gamma_{min} \sim 1.2 - 2.6$),  photon-dominated, low magnetic field dissipation environment. 
Furthermore, the framework allows us to constrain even sub-dominant thermal components that lie below the detector's low energy threshold. Taken together, our results show that optically thin IC scattering offers a physically consistent and observationally viable explanation for the prompt emission 
in a subset of bright GRBs, motivating the application of IC models in future GRB studies.

\end{abstract}



\section{Introduction} \label{sec:intro}
Gamma-ray bursts (GRBs) are among the most powerful explosive transients in the Universe, yet the nature of the radiation processes giving rise to the gamma rays in their relativistic jets remains unresolved \citep{Meszaros2006,Kumar_Zhang2015,Iyyani_2018}. The leading theoretical candidates invoked to explain their broad spectral shapes are synchrotron emission \citep{Rees_Meszaros1994,Tavani1996,Sari1998,Burgess2014a,Macera_etal_2025} and Inverse Compton scattering \citep{Panaitescu_Meszaros2000,Stern2004,Peer&Waxman2004,Nakar_etal_2009,Ahlgren_etal_2015,Iyyani_etal_2015}. Identifying the correct radiation 
model is crucial as they are deeply connected to both the microphysics of particle acceleration and the macrophysical jet dynamics.  

While synchrotron emission has long been considered the leading candidate, its application 
faces significant challenges. In particular, the spectral slopes ($\alpha>-0.67$, \citealt{Preece1998}) and the narrow spectral peak distributions \citep{Goldstein2013,GBMcatalog2014} which are often 
inconsistent with the expectations from standard synchrotron cooling scenarios \citep{Sari1998}. In addition, the GRB spectra tend to exhibit much narrower 
spectral widths  \citep{Axelsson2015} suggesting the spectrum is more consistent with slow cooling rather than fast cooling \citep{Burgess2014a}. However, this leads to another 
difficulty pointing to efficiency where slow-cooling synchrotron requires that most electrons do not radiate efficiently, inconsistent with the 
high radiative efficiencies of $20–90\%$ inferred from observations \citep{Racusin_etal_2009}. Moreover, the extremely high electron Lorentz factors ($\gamma_e \ge 10^4$) 
often implied are difficult to reconcile with acceleration in internal shocks \citep{Iyyani2016,Oganesyan_etal_2019}. Furthermore, implies that only a small fraction of electrons are accelerated to high energies. 

To address these tensions, several modifications to the standard synchrotron picture have been 
proposed. Models with decaying magnetic fields allow electrons to emit in weaker fields, effectively making the spectrum harder at low energies as electrons move from fast to slow 
cooling synchrotron \citep{Pe'er_Zhang2006,Uhm_Zhang2014,Zhang_etal_2016}. Marginally fast-cooling synchrotron considers fine-tuned conditions where the cooling electron Lorentz factor is comparable 
to the injection electron Lorentz factor, producing harder slopes \citep{Daigne2011}. Re-acceleration or slow-heating mechanisms posit that electrons are 
stochastically re-energised before cooling, mimicking a harder effective distribution \citep{Ghisellini&Celotti1999,Kumar&McMahon2008}. In magnetically dominated jets, reconnection 
scenarios can sustain such processes more naturally than baryonic shocks \citep{Beniamini_Piran2014}. Additional alternatives include external shocks to explain 
very high inferred Lorentz factors, and synchrotron self-Compton (SSC) contributions to account for high-energy components \citep{Meszaros_Rees1993,Panaitescu_Meszaros2000,Nakar_etal_2009}. Even hadronic 
synchrotron models have been considered, though their efficiency remains problematic \citep{Waxman1995,Bottcher_Dermer1998,Zhang_etal_2018}. Despite 
these refinements, synchrotron-based scenario is yet to provide a fully self-consistent explanation of the observed spectra across the GRB population \citep{Pe'er2015}. This motivates the search for alternative or complementary radiation channels such as optically thin inverse Compton scattering or photospheric emission dominated models. 

Empirical models such as the Band function, or their extensions through combinations with additional or multiplicative components (e.g. Band + blackbody, Band $\times$ Highecutoff etc), are most commonly employed to characterise GRB prompt spectra \citep{Guiriec2011,Axelsson2012,Iyyani2013,Iyyani_etal_2015,Vianello2018,Sharma_etal_2019,Bordoloi_Iyyani_2025}. These functions are designed to capture the observed spectral shapes with minimal 
bias toward any specific radiation mechanism. In practice, physical interpretations of the underlying radiation processes are often inferred indirectly from such empirical characterisations. 
While this approach provides useful descriptive fits allowing for an informed assessment of the likely underlying radiation
mechanism, it does not fully establish the physical viability of the proposed radiation model. A more 
direct method is to fit the observational data using the radiation models themselves and assess a self-consistent physical scenario. Such implementations already exist for optically thin synchrotron models \citep{Burgess2014a,Zhang_etal_2016,Oganesyan_etal_2019,Burgess_etal_2020,Ryde_etal_2022} and for localised sub-photospheric dissipation models \citep{Ahlgren_etal_2015,Ahlgren_etal_2019,Ahlgren_etal_2022yCat}.

In this work, we focus on the scenario of quasi-thermal Comptonisation in optically thin scenario wherein thermal photons advected from the photosphere propagate to the 
dissipation region, an optically thin zone, where accelerated electrons upscatter them to higher energies. This physical process can produce complex spectral shapes, often featuring 
multiple peaks and breaks, depending on the microphysical properties of the electron distribution, the temperature of the seed thermal photons and optical depth at the dissipation 
site. Such complex spectral shape has recently been identified in GRB 131014A, providing a physically consistent interpretation within this framework 
\citep{Bordoloi_Iyyani_2025}.  

Building on this result, our earlier analysis of GRB 131014A showed that empirical modelling alone already pointed to features characteristic of optically thin inverse Compton scattering \citep{Bordoloi_Iyyani_2025}. However, the physical picture suggested by empirical functions can only be validated through direct fitting with 
a physical model. Motivated by this, the present study extends the investigation to a larger sample of GRBs that display similar spectral characteristics. Specifically, we test the 
scenario in which the prompt emission originates from inverse Compton scattering in the optically thin region above the photosphere, where the bulk kinetic energy is dissipated. 
Our aim is to assess the plausibility of this framework and its potential to yield a consistent physical interpretation of the data.

The paper is organised as follows. In Section 2, we present the selection criteria and the final list of GRBs used for the analysis. Section 3 
describes the optically thin Inverse Compton Scattering (ICS) model, its fit parameters, and the parameter space explored. Section 4 presents the results of the spectral 
fits along with the temporal evolution of the model parameters. In Section 5, we discuss the implications for the microphysics of the dissipation site, the broader physical scenario, 
and the uncertainties associated with the fits. Finally, Section 6 summarises our findings and conclusions.

\section{Criteria \& Sample selection} \label{Sample}
\label{criteria}
We compiled a sample of \textit{Fermi} gamma-ray bursts from the \textit{Fermi} Spectral Catalog\footnote{\url{https://heasarc.gsfc.nasa.gov/w3browse/fermi/fermigbrst.html}} up to 30 
September 2025. For the initial selection, we required the time integrated spectrum to have a fluence $>10^{-5}\,\mathrm{erg\,cm^{-2}}$, a 
low-energy index $+1\ge\alpha > -0.5$\footnote{Since the time-integrated spectrum represents a temporal average of the 
evolving emission, we allowed a slightly softer lower-limit cut on $\alpha$. }, and a high-energy index $-1.7 > \beta > -3.3$. For the corresponding peak flux spectrum of each burst, we additionally required $+1\ge\alpha 
> -0.45$ and $-1.7 > \beta > -3.3$. These criteria yielded a sample of 41 GRBs. Basic properties of this sample, including fluence, $T_{90}$, and the Band-function 
parameters ($E_{\rm peak}$, $\alpha$, and $\beta$) for both the time-integrated and peak spectra are summarized in the Appendix \ref{comparison} in comparison to the {\it Fermi} spectral catalog.

Each GRB in this sample was then subjected to a time-resolved analysis. The time bins were defined using the Bayesian-block algorithm \citep{Scargle2013} applied to the brightest of NaI and 
BGO detectors. Within each bin, we performed spectral fitting with the Band function. We found, however, that during the early-rise and late-decay phases the low signal to noise ratios produced broader Bayesian-block intervals. These broader bins 
occasionally yielded artificially soft $\alpha$ values due to time averaging, and in some cases provided poorly constrained spectral parameters.

To mitigate these effects, we restricted the analysis to the brighter portion of each burst. This interval was defined as the $T_{70}$ region, where $T_{70}$ is 
the time during which $70\%$ of the total background-subtracted counts are accumulated. The start and stop times of this interval 
were determined from the cumulative counts curve of the background-subtracted light curve of the brightest detector, using $15\%$ and $85\%$ levels of the total counts as boundaries.

For each GRB, we then examined whether the majority of the time-resolved bins within the $T_{70}$ interval satisfy the same criteria applied to the peak 
spectrum. If at least $\sim80\%$ of the bins in the $T_{70}$ region met these conditions, the GRB was deemed suitable for direct physical modeling using the Inverse-Compton scenario. 

We additionally tested two alternative spectral models: Band + blackbody and a power law with an exponential cutoff. Three GRBs (GRB 131014A, GRB 200412B, and GRB 
200829A) exhibited clear evidence for an additional blackbody component when fit with the Band + blackbody 
model. In some cases, individual time-resolved bins were better described by a cutoff power law than by 
either the Band or Band + BB models (e.g. GRB100805845, GRB231129799 etc). GRBs showing such behavior were also excluded from the final sample.

\begin{figure*}
    \includegraphics[width=1.0\linewidth]{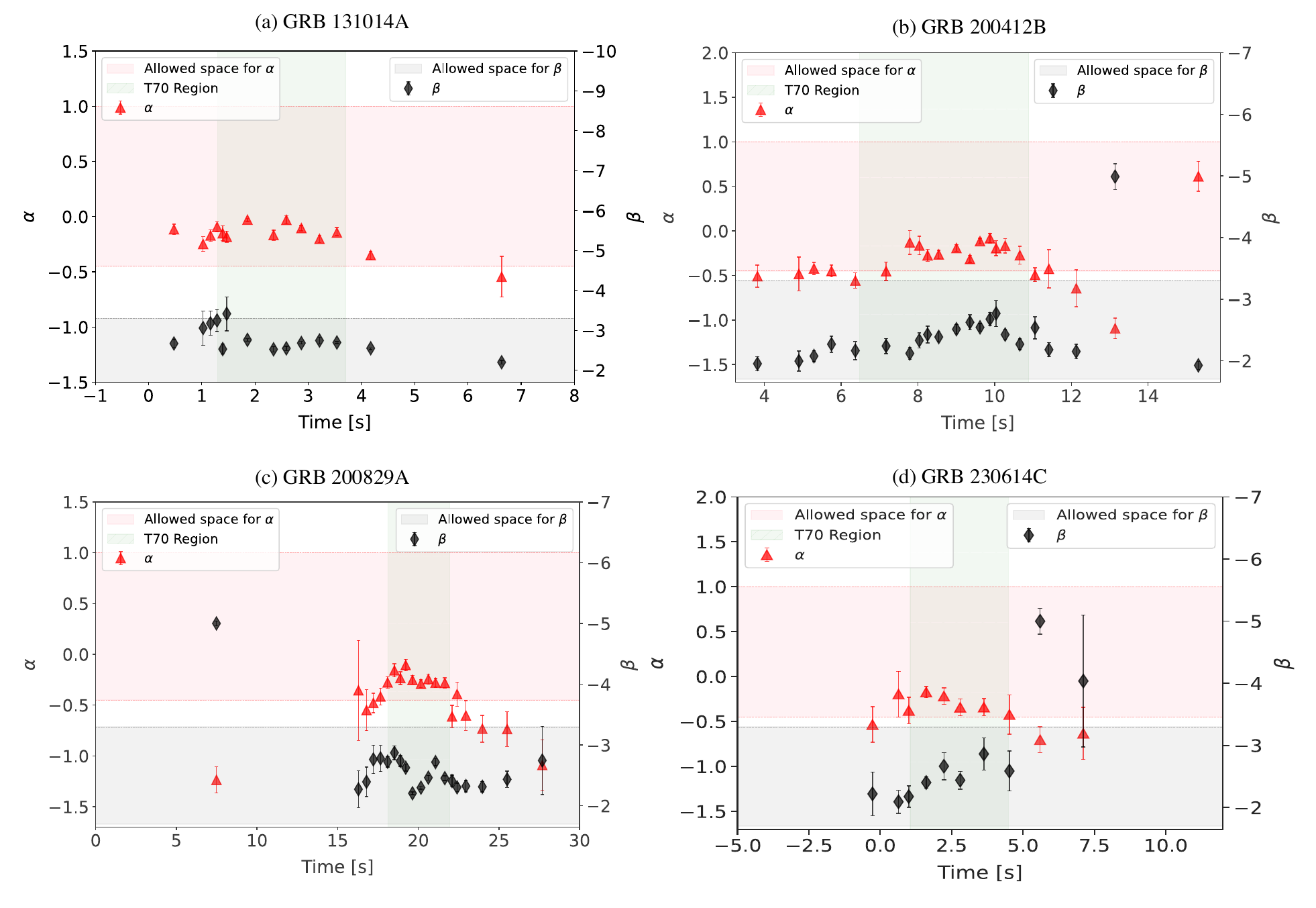}
    \caption{The temporal evolution of the Band $\alpha$ (red triangles) and $\beta$ (black diamonds) within the $T_{70}$ interval (light-green vertical band) for the selected GRBs are shown. The red and grey horizontal shaded regions indicate the allowed ranges of $\alpha$ and $\beta$ used in the sample selection.}
    \label{Band_alpha_beta_accepted_GRB}
\end{figure*}

Applying this procedure reduced the final sample to four GRBs that satisfy the selection criteria both in their time-integrated spectra and across the time-resolved bins within the $T_{70}$ region (Figure \ref{Band_alpha_beta_accepted_GRB}). These bursts are GRB 131014A (GRB131014215), GRB 200412B (GRB200412381), GRB 200829A (GRB200829582), and GRB 230614C (GRB230614424). 

GRB 131014A was detected by several high-energy observatories, 
including the {\it Fermi} GBM and LAT 
\citep{Fermi_GBM_131014A,Fermi_LAT_131014A}, Konus-Wind \citep{Konuswind_131014A}, and the Suzaku/WAM instrument \citep{Suzaku_131014A}. The burst was localized at RA = 
$100.29^{\circ}$ and Dec = $-19.13^{\circ}$, with an uncertainty of about $53'$ \citep{IPN_131014A}. Approximately 12 hours after the trigger, Swift-XRT Target-of-Opportunity observations identified a 
fading X-ray source \citep{Swift_XRT_13104A}, and subsequent reports noted a possible associated optical transient \citep{NOT_OT_131014A,GROND_OT_131014A,Swift_OT_upperlimit_131014A}. However, no secure redshift measurement was obtained for this event.

GRB 200412B was detected by {\it Fermi} \citep{200412B_Fermi_GBM,200412B_Fermi_LAT}, 
Konus-Wind \citep{200412B_Konus_Wind}, CALET \citep{200412B_CALET}, and AstroSat \citep{200412B_AstroSat}. The burst was localized at 
RA = $277.48^{\circ}$ and Dec = $61.79^{\circ}$ with an uncertainty of $0.5^{\circ}$. An optical afterglow was identified by multiple facilities \citep{200412B_Optical_MASTER,200412B_Optical_DOT,200412B_Optical_Mondy_Terskol,200412B_Optical_TNT}. Despite extensive follow-up observations, no redshift measurement has been reported for GRB 200412B.

GRB 200829A was detected by {\it Fermi} \citep{200829A_Fermi_GBM}, Konus-Wind \citep{200829A_Konus_Wind}, and the {\it Swift}/Burst Alert Telescope (BAT; \citealt{200829A_Swift_BAT}). The burst was 
localized at RA $=251.13^{\circ}$ and Dec $=72.36^{\circ}$. An optical afterglow was subsequently identified by multiple 
observatories \citep{200829A_Optical_MASTER,200829A_Optical_Swift}. The multi-band afterglow light curve of GRB 200829A is found to exhibit a clear chromatic evolution, featuring a plateau phase that gradually transitions into a power-law decay \citep{200829A_afterglow_2023}. A redshift of $z = 1.25 \pm 0.02$ was measured through {\it Swift}-UVOT photometry \citep{200829A_redshift}. 

GRB 230614C was detected by {\it Fermi} \citep{230614C_Fermi_GBM}, AstroSat \citep{230614C_AstroSat} and GRBAlpha \citep{230614C_GRBAlpha}. The burst was localised at RA = $229.21^{\circ}$ and Dec $= 10.30^{\circ}$. There is no afterglow observations reported for this burst. 

The observational properties including trigger time (Universal Time, UT), localisation (RA, Dec), energy fluence in $10-1000$ 
keV and $T_{90}$ estimated in $50-300$ keV, of these four GRBs are listed in Table~\ref{tab:GRB_summary}. The detailed time-resolved spectral results of the Band function fits for all 41 initially selected GRBs, 
including their full sets of spectral evolution plots of $\alpha$ and $\beta$, are provided as supplementary material. A small number of representative cases are shown in the Appendix \ref{sample}.

\subsection{Motivation for the Criteria}
The above mentioned criteria are designed to isolate brighter bursts whose instantaneous spectra are consistent with the presence of a strong thermal component at low energies and an electron driven 
high-energy tail. A high-energy fluence greater than $10^{-5}\,\rm erg \,cm^{-2}$ ensures the selection of bright GRBs, providing higher signal-to-noise ratios and enabling finer time resolution in the time-resolved spectral analysis.

A relatively hard low-energy index ($\alpha > -0.45$) selects cases in which the unscattered thermal seed photons remain within the observed energy window. This choice is motivated by the results of \citet{Acuner_etal_2019}, who demonstrated that synthetic spectra from non-dissipative photospheric 
emission, when folded through the {\it Fermi} Gamma ray Burst Monitor (GBM) detector response and fit with empirical function like Band function, typically yield $\alpha$ values in the range $-0.4$ to $0.0$. Although 
softer $\alpha$ values can be expected when the observed emission is dominated by 
Comptonised rather than seed photons supposedly when the seed is outside the observation window \citep{Ahlgren_etal_2019}, such cases become degenerate 
with optically thin synchrotron spectra. To avoid this ambiguity, we restrict our analysis to GRBs with sufficiently hard low-energy slopes to serve as clear candidates for quasi-thermal seed Comptonisation.

Diffusive shock-acceleration theory generally predicts electron power-law indices $p \approx 2$--$2.2$ for parallel non-relativistic or ultra-relativistic shocks 
\citep{kirk_1987, kirk_2000}. However, observational studies such as \citet{Shen_etal_2006} show that $p$ can span a broader range ($\sim$1.6--4) across GRB prompt emission, X-ray 
afterglows, blazars, and pulsar wind nebulae. The requirement $-1.7 > \beta > -3.3$ ensures compatibility with expectations 
for electron acceleration in optically thin regions. Dissipation at modest optical depths ($\tau <1$) is expected to generate a post-shock electron distribution consisting of a 
thermal population plus a non-thermal power-law tail. In the optically thin limit, considering higher-order scatterings are 
suppressed, the Comptonised spectrum is dominated by first-order scattered photons and therefore closely reflects the 
shape of the underlying electron energy distribution. Our adopted $\beta$ threshold therefore conservatively excludes 
spectra whose steepness would be inconsistent with optically thin Comptonisation or would imply substantial thermalisation at sub-photospheric depths.

Taken together, these physically motivated criteria select bright GRBs that are the most promising candidates for direct physical modeling within an optically thin quasi-thermal Comptonisation framework.

\begin{table}[h!]
\centering
\small
\begin{tabular}{l c c c c c c l}
\hline
\textbf{GRB Name} & \textbf{RA } & \textbf{DEC } & 
\textbf{Trigger Time } & \textbf{Fluence } & 
\textbf{T$_{90}$} & \textbf{Redshift} & 
\textbf{Detectors Used} \\
& \textbf{(deg)} & \textbf{(deg)} & \textbf{(UTC)} & \textbf{(erg cm$^{-2}$)} & \textbf{(s)} & & \\
\hline
\textbf{GRB131014A} & $100.29$ & $-19.13$ &
2013-10-14 05:09:00 & $1.98 \pm 0.002\times10^{-4}$ & 3.20 $\pm$ 0.09 &
--- & NaI(9,10,11), BGO(1),\\
& & & & & & & LLE, LAT \\
\textbf{GRB200412B} & $277.48$ & $61.79$ &
2020-04-12 09:08:41 & $7.25 \pm 0.04\times10^{-5}$ & 6.08 $\pm$ 0.29 &
--- & NaI(6,7,8), BGO(1) \\
\textbf{GRB200829A} & $251.13$ & $72.36$ &
2020-08-29 13:58:15 & $2.14 \pm 0.001\times10^{-4}$ & 6.90 $\pm$ 0.36 &
1.25 $\pm$ 0.02 & NaI(4,8), BGO(0,1) \\
\textbf{GRB230614C} & $229.21$ & $10.30$ &
2023-06-14 10:10:31 & $2.01 \pm 0.007\times10^{-5}$ & 6.08 $\pm$ 0.36 &
--- & NaI(9,10,11), BGO(1) \\
\hline
\end{tabular}
\caption{The observational details of the selected GRBs are listed above.}
\label{tab:GRB_summary}
\end{table}

\begin{figure*}[!ht]
    \centering
    \includegraphics[width=\linewidth]{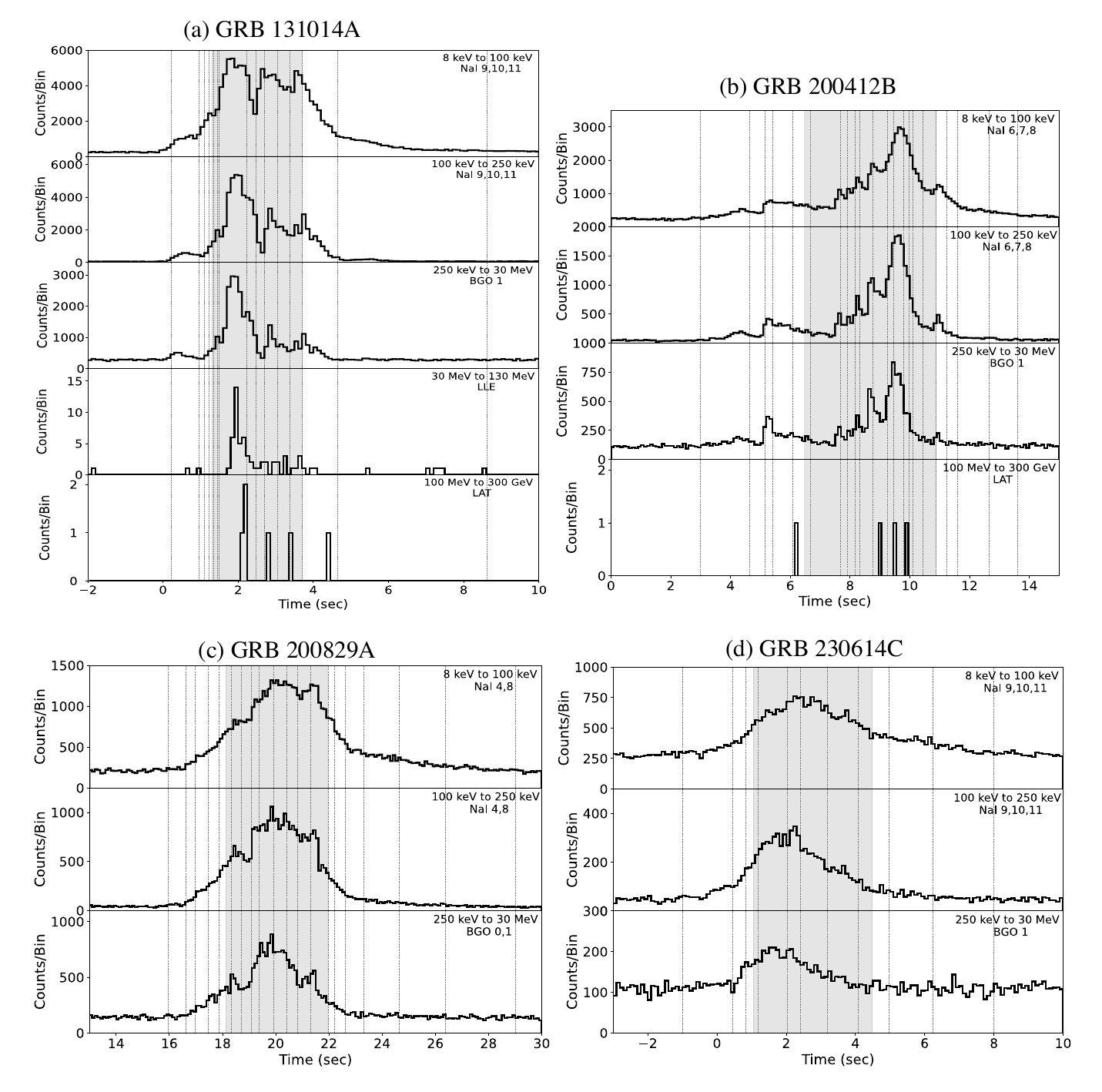}     
    \caption{The multipanel figures show the counts per bin light curves for $0.5$ second binning, resolved into three 
    energy ranges: 8-100 keV, 100-250 keV, and 250 keV-30 MeV for (a) GRB 131014A, (b) GRB 200412B, (c) GRB 200829A, and 
    (d) GRB 230614C. For GRB 131014A, the light curves from LAT-LLE (30-100 MeV) and LAT ($>100$ MeV) are also included. The vertical dashed lines indicate the time resolved intervals 
    determined using Bayesian Block binning applied to the time integrated segment of the burst, while the shaded grey region represents the $T_{70}$ interval of the burst.}
    \label{lightcurves}
\end{figure*}

\section{Optically thin inverse Compton scattering model} 

In the baryonic fireball framework, the prompt gamma-ray emission comprises two principal components: 
a thermal component released when the outflow becomes transparent at the photosphere forming in the coasting phase, and a non-thermal 
component produced in the optically thin regions above it.
In this picture, the outflow is advected through the photosphere, and any temporal variation in the flow properties naturally leads to corresponding variations in the observed 
photospheric emission. These properties are governed by the initial conditions at the central engine, such as the burst luminosity $L(t)$, the dimensionless entropy $\eta(t) \equiv 
\Gamma(t) = L(t)/\dot{M}c^2$, and the nozzle radius $R_0(t)$, where $\dot{M}$ denotes the baryon loading and $t$ is time, without assuming any particular parametric prescription for 
their temporal evolution.  We do not impose any specific parametric prescription on their temporal evolution.  We further assume that the jet dynamics are predominantly adiabatic, following the classical fireball evolution.
As the jet expands beyond the photospheric 
radius, part of its kinetic energy can be dissipated through 
internal shocks, collisional heating, or plasma instabilities. At the site of these dissipation processes, 
the electrons get accelerated to relativistic energies, creating a population capable of producing high-energy radiation.

In regions where the magnetic field is insufficient for synchrotron cooling to dominate, the 
accelerated electrons primarily lose energy by upscattering the 
soft thermal photons advected from the photosphere. This interaction produces a non-thermal spectrum 
through optically thin inverse Compton scattering (ICS). Because the medium is transparent, photons 
typically undergo only a few scatterings before escaping, causing the emergent spectrum to closely 
track the underlying electron energy distribution.

For this work, we adopt the ICS framework developed in \citealt{Bordoloi_Iyyani_2025} to perform direct 
physical model testing on our GRB sample. The post-shock electron distribution is described as a 
combination of a thermal population and a non-thermal 
power-law tail. The seed photon field is taken to be the thermal component carried outward 
from the photosphere to the dissipation region.

The accelerated electrons in the shock layer cool through ICS on co-moving timescales given by
\begin{equation}
    t'_{\rm cool} \simeq \frac{3 m_e c}{4\,\gamma_e\,\sigma_T\,u_{\rm ph}} 
\end{equation}
\noindent
where $\sigma_T$ is the Thomson cross-section, $c$ is the speed of light, $m_e$
is the electron mass, $\gamma_e$  is the electron Lorentz factor, and $u_{\rm ph}$ is the radiation energy 
density of the seed thermal photons at the dissipation site. For typical GRB parameters, these 
cooling timescales are on the order of a few microseconds or less. By contrast, considering the bulk Lorentz factor of a few hundreds, the observed time-resolved 
spectra are accumulated over intervals 
ranging from a fraction of a second to several seconds. The spectra we model are therefore intrinsically 
time-averaged representations of rapidly evolving emission.

To account for this temporal averaging, within the {\it Naima} framework \citep{naima}, we adopt an empirical exponentially 
cutoff power-law function to describe the seed photon distribution rather than a pure Planck spectrum. This approach 
captures the effective spectral shape of the 
photon field that participates in the ICS process during the longer observational integration times.

\subsection{{\it Naima} ICS Model Setup}
\label{naima_model}
{\it Naima} computes inverse Compton emission by modeling how a population of relativistic electrons scatters soft 
photons to higher energies \citep{IC_Naima_model_2014}. The implementation 
is physically grounded and numerically robust, and is widely used for modeling non-thermal radiation in 
high-energy astrophysics. 
{\it Naima} does not assume a fixed shape for the electron distribution or the seed component.  
{\it Naima} uses the full Klein–Nishina cross section ensuring accurate modelling at higher 
electron Lorentz factors, high seed photon energies thereby properly accounting for Klein–Nishina 
suppression, spectral curvature at high energies and shift in scattering peaks. \\
\\
\noindent
{\bf Seed Photon Distribution in the Co-Moving Frame}\\
The soft photon field consists of a cutoff power-law spectrum that represents thermal radiation 
advected from the photosphere. In the co-moving frame, its differential energy density 
(energy per unit volume) is defined as

\begin{equation}
u_{\rm ph,seed}(E')
=
4\,\frac{K}{c} \left( \frac{E'}{E_{0}} \right)^{2 + \alpha_{seed}}
\exp\!\left( -\frac{E'}{E_{\rm c,seed}} \right).
\end{equation}
\noindent
where $K$ is the normalization of the unit $\rm keV cm^{-2} s^{-1}$, $\alpha_{seed}$ represents the power law index, $E_0$ is a reference energy and $E_{c,seed}$ is the cutoff energy.
\\
{\bf Electron Distribution in the Co-moving Frame}
\noindent
\\
The electron population is represented by a thermal Maxwell–Jüttner core with a non-thermal power-law tail:
\begin{equation}
    N_{e}(\gamma_e)
=
N_0\left[
\frac{\gamma_e^{2}\sqrt{1-\gamma_e^{-2}} 
\exp(-\gamma_e/\theta)}
{\theta K_{2}(1/\theta)}
+
\epsilon\left(\frac{\gamma_e}{\theta}\right)^{-\delta}
\Theta\!\left(\frac{\gamma_e}{\gamma_{\min}}\right)
\right],
\end{equation}
\noindent 
where $N_0$ is the normalization, $\gamma_e = \frac{E}{m_{e}c^{2}}$
is the electron Lorentz factor, $\theta = \frac{kT_e} {m_{e}c^{2}}$ is the dimensionless 
temperature, $\delta$ is the index of the non-thermal 
tail, $\gamma_{\min} = 1 + \kappa \theta$ sets the transition from the thermal core to the tail, $\kappa = 3$ \citep{Baring_1993,Baring_Braby2004,Burgess2014a} 
$\epsilon$ ensures continuity between the components (refer \citealt{Bordoloi_Iyyani_2025} for more details), 
$\Theta$ is the Heaviside step function, and $K_{2}$ is the modified Bessel function of the 
second kind. Please note here $N_{e}(\gamma_e)$ is total electron number distribution per $\gamma_e$ and not number density. 
\\
\noindent
{\bf IC Emission Calculation in {\it Naima} in the Co-moving Frame}\\
\noindent
{\it Naima} computes the IC spectrum by convolving the seed photon density with the 
electron distribution using the full Klein–Nishina cross section:
\begin{equation}
    \frac{{\rm d}N'}{{\rm d}t' {\rm d}E'_{IC}} 
=
\int N_{e}(\gamma_e)
\int \frac{u_{\rm ph,seed}(E')}{E'}\,
c\,
\frac{{\rm d}\sigma_{\rm KN}}{{\rm d}E'_{IC}}
\,{\rm d}E'\,{\rm d}\gamma_e.
\label{IC_equation}
\end{equation}
\noindent 
where $E'_{IC}$ represents the IC scattered photon energy while $E'$ represents the 
seed photon energy before scattering. The inner integral computes the IC 
output from one electron at energy $\gamma_e$. It considers all possible seed photon 
energies $E'$ weighs them by how efficiently they scatter into 
$E'_{IC}$. This term can be summarised as the $P(E'_{IC},\gamma_e)$ which is the single electron IC 
emissivity. The outer integral then sums the contributions from all electrons in the 
electron distribution. The above convolution yields photon production per unit time and energy 
in the co–moving frame. 
Multiplying by $E'_{\rm IC}$ gives the luminosity per unit energy, 
\begin{equation}
    L'_{\rm IC}(E'_{\rm IC})
    =
    E'_{\rm IC}
    \frac{{\rm d}N'}{{\rm d}t'\,{\rm d}E'_{\rm IC}}.
\end{equation}
Since inverse Compton cooling occurs on microsecond scales, far shorter than GBM time 
bins. Thus, the electron and seed-photon distributions supplied to the IC kernel represent 
time-averaged states within each bin, and the IC 
emissivity is evaluated as an instantaneous steady state solution appropriate for that interval. Thus, equation \ref{IC_equation} represents the standard isotropic IC kernel used by {\it Naima}, evaluated here with user defined seed and electron distributions in the co-moving frame. 
\\
\noindent
{\bf Transformation to the Observer Frame}\\
The mapping between co–moving and observer–frame quantities includes both
relativistic boosting and cosmological redshift:
\begin{equation}
    t_{\rm obs} = \frac{1+z}{\mathcal{D}}\,t', \qquad
    E_{\rm obs} = \frac{\mathcal{D}}{1+z}\,E'_{\rm IC}, \qquad
    \label{eq:transformations}
\end{equation}
where
\begin{equation}
    \mathcal{D}
    =
    \frac{1}{\Gamma(1-\beta\cos\theta)}
\end{equation}
is the Doppler factor, $\theta$ is direction of the outflow velocity vector ($\beta$) with respect to the 
line of sight, and $z$ is the redshift. The transformation follows:
\begin{equation}
    \frac{{\rm d}N}{{\rm d}t_{\rm obs}\,{\rm d}E_{\rm obs}}
    =
    \frac{{\rm d}N'}{{\rm d}t'\,{\rm d}E'_{IC}}.
\end{equation}
{\it Naima} computes the IC emissivity assuming isotropy in the co-moving frame and the Lorentz Doppler factors associated with the transformations of energy 
and time cancel when the emission is integrated over the relativistically beamed solid angle. In other words, for on-axis viewing, although relativistic aberration confines the observed emission to a 
narrow cone of opening angle $\sim 1/\Gamma$ across the line of sight, the solid-angle integration of the Doppler-boosted emissivity (photons per second per keV per solid angle) over this beamed region, 
recovers the same photon number spectrum per unit time and energy as in the co-moving frame. Consequently, for an on-axis ultra-relativistic outflow, the photon number spectrum 
per unit time and energy remains invariant between the co-moving and observer frames.
\\
{\bf Observer–Frame Photon Flux}
\noindent
\\
The differential photon flux at Earth follows from geometric dilution via the
luminosity distance $d_{L}$:
\begin{equation}
    N_{\rm obs}(E_{\rm obs})
    =
    \frac{1}{4\pi d_{L}^{2}}\,
    \left.
        \frac{L'_{\rm IC}(E'_{\rm IC})}{E'_{\rm IC}}
    \right|_{E'_{\rm IC} = \frac{1+z}{\mathcal{D}}E_{\rm obs}} ,
    \label{eq:final_flux}
\end{equation}
with final units of ${\rm photons~cm^{-2}~s^{-1}~keV^{-1}}$.
This is the input photon spectrum for the forward-folding process which when convolved through the instrument 
response produces the predicted model counts 
\begin{equation}
C^{\rm model}_{i}
=
\int_{0}^{\infty}
R_{i}(E)\,
N_{\rm obs}(E)\,
{\rm d}E,
\end{equation}
\noindent
where $C^{\rm model}_{i}$ is the predicted model counts per channel $i$ and $R_{i}(E)$ is the detector response 
matrix. Those are compared with the $C^{\rm data}_{i}$
which are the observed {\it Fermi} detector counts per channel. 

The free parameters of the model include the electron distribution quantities $N_0$, $\delta$, and the electron thermal 
energy $E_{\rm th} \equiv kT_e$ where $T_e$ is the electron temperature, together with the 
seed photon distribution parameters $K$, $\alpha_{seed}$, and $E_{c,\rm seed}$, as well as the bulk Lorentz 
factor $\Gamma$. Please note that both the electron thermal energy, $E_{\rm th}$, and the cutoff of the seed photon 
distribution, $E_{c,\rm seed}$ values reported are in the co-moving frame of the outflow unless otherwise mentioned.
The luminosity distance $d_L$ is fixed to $6.8 \times 10^{6} \,\mathrm{kpc}$, corresponding to a redshift of $z=1$ for 
GRB 131014A, GRB 200412B, and GRB 230614C. For GRB 200829A, where the redshift, $z=1.25$ has been measured, we use the 
corresponding luminosity distance $d_L = 8.96 \times 10^6 \, \rm kpc$. All distances are computed assuming a standard $\Lambda$CDM cosmology with 
$H_0 = 67.4 \pm 0.5~\mathrm{km\,s^{-1}\,Mpc^{-1}}$, $\Omega_m = 0.315$, and $\Omega_{\Lambda} = 0.685$ \citep{Planck_2020}.

Furthermore, in the {\it Naima} formalism, it is worth noting that the inverse Compton calculation assumes that 
the entire seed photon distribution is available for 
scattering with the electrons in the co-moving frame (equation \ref{IC_equation}). However, in our spectral 
fits the observed spectrum generally contains both the unscattered thermal component 
and the Compton scattered component, ensured by our requirement that the low energy index satisfy $\alpha>-0.45$. 
Therefore, the fitted seed photon 
distribution should be interpreted as representing the total thermal photon flux available at the dissipation site. 
Only a fraction of these thermal photons undergo 
inverse Compton scattering to produce the non-thermal high-energy component of the spectrum, while the remainder 
appears directly as the unscattered thermal emission (also refer to section \ref{thermal_constraint}).


\subsection{Fit Parameter Space}
\label{param_space}
In this physical model of optically thin inverse Compton scattering the 
thermal radiation is advected from the photosphere which forms in the coasting phase (i.e above the saturation radius) serves as the seed photon field. After decoupling at the 
photosphere, this thermal emission undergoes only geometric (inverse-square) dilution without any 
modification to its spectral shape or temperature. Upon reaching the dissipation region, these photons are 
upscattered by the heated or accelerated electrons, 
producing the observed IC component. In the {\it Naima} framework, only a single scattering order is included. 
Given that the dissipation zone is optically thin and the photon–electron interaction probability is 
correspondingly small, restricting the computation to first-order scattering constitutes a reasonable and 
physically justified approximation for this modelling approach. 

This physical scenario naturally fits within the classical fireball framework. Using the burst flux ($F_{ob}$) 
derived for each time interval during the empirical 
fitting stage (carried out as part of the selection procedure; see Section~\ref{criteria}), and adopting either 
the measured or an assumed redshift for the burst, we estimate the isotropic luminosity ($L = 4\pi d_L^2 \rm YF_{ob}$), where $Y$ is the inverse of radiation efficiency\footnote{In line with the results of \citet{Racusin2011}, we adopt an average radiative 
efficiency of $0.2$ for GRBs detected only by the \textit{Fermi} GBM, while for GRBs with LAT detections we assume a higher average efficiency of $0.65$.}, for each interval of the prompt emission. Together with a 
physically motivated range for the nozzle radius, $R_0 \sim 10^6$–$10^9~\mathrm{cm}$ \citep{Iyyani2013, Iyyani2016}, 
these inputs enable us to place informed 
bounds on model parameters. For this we adopt the methodology of outflow parameters based on the basic fireball 
model framework presented in \citealt{Pe'er2007}. \\

The seed cutoff energy $E_{c,\rm 
seed} = 3.9kT'(R_{ph})/(\alpha_{seed}+2)$, where $k$ is the Boltzmann constant, is obtained using the expression of co-moving photospheric temperature
\begin{equation}
T'(R_{\rm ph}) 
= 
\left( \frac{L}{4\pi R_{0}^{2} c a} \right)^{1/4}
\, \Gamma^{-1}
\left( \frac{R_{\rm ph}}{R_{s}} \right)^{-2/3} ,
\label{comoving_Trph}
\end{equation}
\noindent
where $a$ is the radiation constant, $R_{ph} = (L\,\sigma_T/8\pi\Gamma^3 m_pc^3)$ is the photospheric radius, $R_s = \Gamma R_0$ is the saturation radius, and considering a range of $\alpha_{seed}$ between $-0.5$ and $+1$. 
The equation \ref{comoving_Trph} reduces to:
\begin{equation}
    \frac{T'(R_{\rm ph})}{\Gamma^{5/3}} \propto 
    \frac{R_0^{1/6}}{L^{5/12}}
    \label{Tph_constrain}
\end{equation}
which shows that for the given range of luminosity $L$ and nozzle 
radius $R_{0}$, the allowed combinations of $\Gamma$ and $T'(R_{\rm ph})$ are constrained. In practice, the inferred values of $E_{c,{\rm seed}}$ and $\Gamma$ are
required to satisfy the relation in equation~\ref{Tph_constrain}, while also producing statistically acceptable spectral fits with random,
structure–free residuals.

The normalization of the electron distribution, $N_0$, is restricted using an order-of-magnitude estimate of the total electron content implied by the burst 
luminosity and the allowed $\Gamma$ range using the below formula:
\begin{equation}
N_{0} \sim \frac{L}{\Gamma\, m_{p} c^{2}} \times t_{\rm obs}
\end{equation}
\noindent
where $t_{obs}$ is the duration of each time interval.
The power-law index $\delta$ is explored over a conservative range of 
$1$ to $5$, guided by the results of \citealt{Bordoloi_Iyyani_2025, Shen_etal_2006} and by theoretical expectations \citep{kirk_1987,kirk_2000}, at the same time ensuring that the parameter is 
properly constrained. In contrast, the seed-photon normalization (K) and the electron thermal energy ($E_{th}$) are left free over a broader respective intervals during the fitting.

We further note that previous studies such as \citealt{Iyyani2013,Preece2014,Iyyani2016} have shown that within a classical fireball framework, outflow parameters such as $\Gamma$ 
and $R_0$ can evolve with a particular trend within individual emission pulses, whereas some hybrid outflow models assume these quantities remain constant \citep{Gao_etal_2015,Yan_2024}. The bursts analysed here exhibit strong thermal components indicative of a 
dominant hot fireball jet and are largely multi-pulsed, implying that time-resolved intervals represent effective averages over multiple underlying pulses. Accordingly, we constrain $\Gamma$ and 
$R_0$ within physically motivated ranges based on previous studies, without enforcing temporal linking of these parameters across time intervals, ensuring independent interval fitting while avoiding unphysical parameter space. 

Furthermore, we assume the outflow to be quasi-static within each time interval defined by the Bayesian Block binning. The light curves are segmented using the Bayesian Block algorithm, which adaptively traces the intrinsic temporal variations of the emission; 
hence, the relevant variability timescales correspond to the widths of these blocks. These bin widths are significantly longer than the dynamical time required for the outflow to reach the photosphere (of 
the order of milliseconds), implying that the central engine properties evolve only across successive blocks and remain approximately constant within each interval \citep{Iyyani2013}. We do not assume that a single shell is responsible for the emission in a given bin. Rather, multiple shells or fluid elements may be emitted during this period, but are launched under similar central 
engine conditions, sharing comparable intrinsic properties such as the injection luminosity, bulk Lorentz factor, and nozzle radius. Consequently, the inferred parameters characterize the 
representative state of the outflow during that interval. In the case of multi-pulsed bursts, a given block may still contain unresolved sub-structures or overlapping pulses; therefore, the derived physical parameters should be interpreted as effective, time-averaged quantities describing the collective emission within that interval.

After obtaining a successful fit, we use the best-fit parameters of the seed-photon distribution 
to infer the corresponding outflow properties using the above equations, which are presented in Section \ref{derived_values}.

\section{Spectral Analysis} 
\label{spectral_fit}

The spectral analysis was performed using the Multi-Mission Maximum Likelihood framework (3ML; \citealt{Vianello_etal_2015,3ml}). For each GRB, we selected data from the three 
brightest Sodium Iodide (NaI) detectors with source angles $<60^{\circ}$ \citep{GBMcatalog2014}, together with the brightest Bismuth Germanate (BGO) detector. The specific detectors used for each burst are listed in 
Table \ref{tab:GRB_summary}. For GRB 131014A, the analysis additionally incorporated LAT-Low Energy (LLE) data and LAT photons above 100 MeV. In contrast, GRB 200412B exhibited only a few LAT events above 100 MeV 
\citep{200412B_Fermi_LAT}, insufficient for a meaningful contribution to the likelihood analysis; therefore, LAT was not included for this burst while no LLE data were reported for this event.

The energy ranges adopted for the analysis were as 
follows: 8 - 900 keV for the NaI detectors, 250 keV - 30 MeV for BGO. The data in the interval 30–40 keV
in the NaI detectors are ignored because of the iodine K-edge \citep{Bissaldi_etal_2009}.  For GRB 131014A, 
data in the energy range 30 - 100 MeV for LAT–LLE, and 100~MeV--2~GeV for LAT was utilised. 
The source interval: 0 -- 8.6 s, 3 -- 17 s, -1 -- 30 s, -1 -- 8 s were chosen for the bursts GRB 131014A, GRB 200412B, GRB 200829A and GRB 230614C, 
respectively. The background for each detector was modeled using a polynomial function fitted to count 
data in time intervals preceding and following the source interval. 
Spectral fitting was carried out within a maximum-likelihood estimation (MLE) framework using the Poisson–Gaussian likelihood, which simultaneously accounts for the Poisson nature of the total observed counts and the Gaussian 
uncertainties associated with the background measurements \citep{Burgess_etal_2019}. 
The fit statistic employed was the corresponding negative log-likelihood, 
PGstat\footnote{https://heasarc.gsfc.nasa.gov/xanadu/xspec/manual/XSappendixStatistics.html}. 
The optimisation of PGstat is performed using the MINUIT package, where the 
MIGRAD algorithm determines the best-fit solution. Around the maximum-likelihood 
point, the HESSE routine computes the covariance matrix from the Hessian of the 
negative log-likelihood under the assumption that the likelihood surface is locally 
well approximated by a Gaussian. The diagonal elements of this covariance matrix 
provide the variances of the parameters, while the off-diagonal elements quantify 
their covariances. The Pearson correlation coefficients presented in 
Appendix~\ref{fit_correlations} are derived from this covariance matrix by normalising 
the off-diagonal elements by the product of the corresponding parameter standard 
deviations. It is important to note that the Hessian-based uncertainties rely on the Gaussian approximation, the 
resulting standard deviations can only be interpreted as lower bounds on the true parameter uncertainties in directions where strong parameter degeneracies are 
present. 3ML fit result, however, also reports asymmetric confidence intervals obtained using the MINOS algorithm, which determines parameter bounds by profiling the likelihood and identifying the values where
$\Delta(-2 \ln \mathcal{L}) = 1,$
corresponding to the $68\%$ confidence interval for a single parameter. The parameter uncertainties quoted throughout the text and shown in the figures correspond to these MINOS profile-likelihood intervals, which naturally account for asymmetries in the likelihood surface. Several of the reported parameters therefore exhibit asymmetric error bars.

Since PGstat does not follow a known analytic distribution, we assess the goodness of fit  by a combination of manual 
inspection of the residuals, with acceptable fits exhibiting statistically consistent, structureless 
residual patterns and no systematic deviations, parameters are well constrained, the errors are finite and meaningful. 


For GRB 131014A, which includes LLE and LAT data, we applied the detector effective-area correction factors described in \citealt{Bordoloi_Iyyani_2025}. For the remaining GRBs, where only GBM data were used, no effective-area corrections were required.

\begin{figure*}[!ht]
    \centering
    \includegraphics[width=\linewidth]{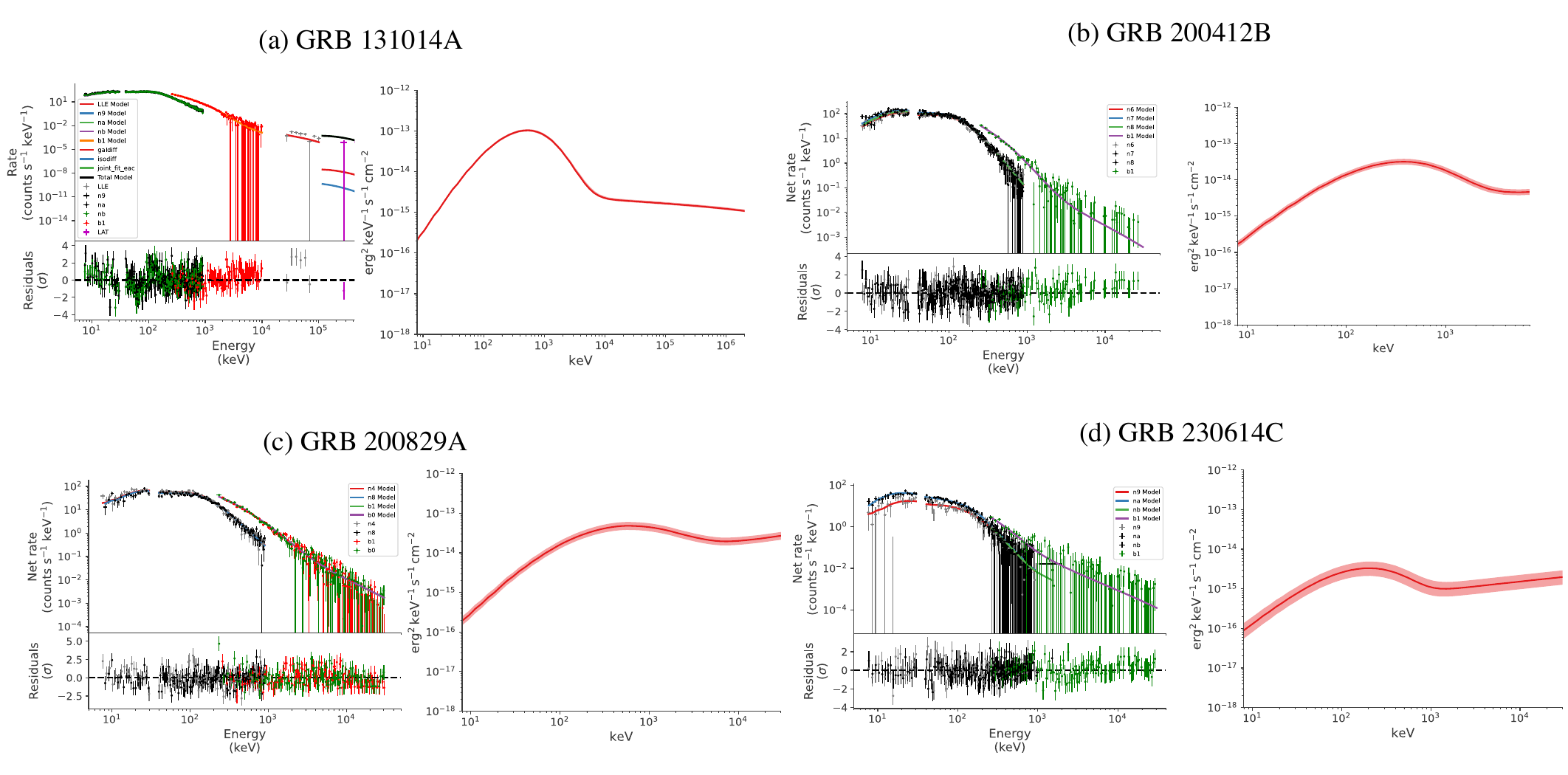}
    \caption{The left panel shows the observed count spectra (top) together with the best-fit model counts (solid curves) and their corresponding residuals (bottom), while the right panel presents the associated $\nu F_{\nu}$ representation of the IC-model fit for the peak time interval of (a) GRB 131014A, (b) GRB 200412B, (c) GRB 200829A, and (d) GRB 230614C. }
    \label{spectral_fits}
\end{figure*}

The optically thin IC model implemented through the {\it Naima} framework (Section \ref{naima_model}) was fitted 
to each time-resolved bin obtained via the Bayesian Block procedure (Section \ref{criteria}) for all GRBs in the sample. 
The counts spectra, associated residuals, and the corresponding $\nu F_{\nu}$ representations of the 
model fit for the peak bin of each burst are shown in Figure \ref{spectral_fits}. The residuals indicate statistically random scatter without any significant structural trends or waviness, confirming the adequacy of the fits.

\begin{figure*}[!ht]
    \centering
    \includegraphics[width=\linewidth]{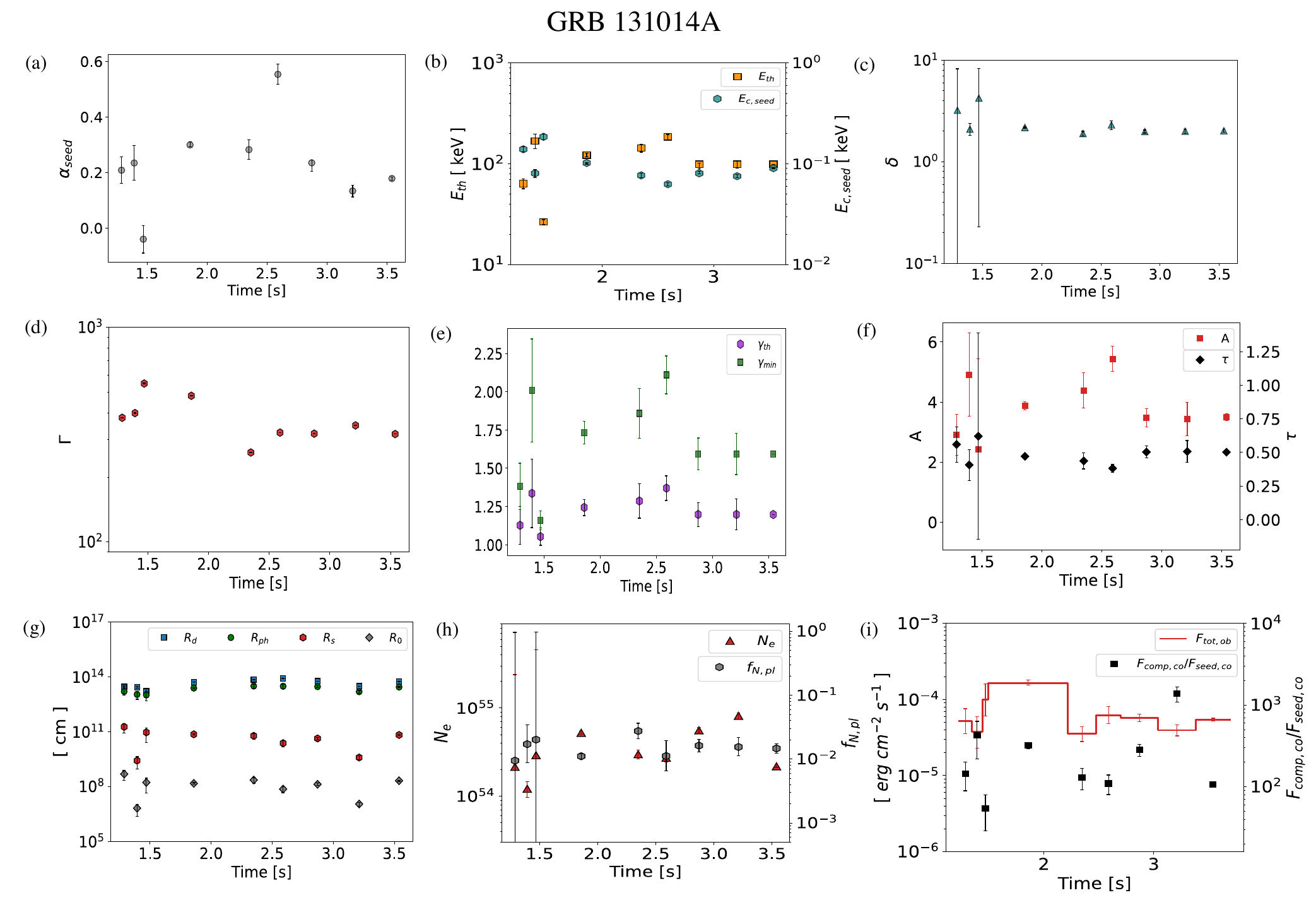}
 \caption{The temporal evolution of the best-fit parameters of the IC model for GRB 131014A is shown for:
  (a) the seed spectral index, $\alpha_{seed}$ (grey circles),
  (b) $E_{\mathrm{th}}$ (orange squares) and $E_{c,\mathrm{seed}}$ (green hexagons),
  (c) $\delta$ (green triangles), and
  (d) $\Gamma$ (red hexagons). The temporal evolution of the derived physical parameters of the GRB131014A: (e) $\gamma_{th}$ (purple hexagon) and $\gamma_{min}$ (green square), (f) 
    $A$ (red squares) and $\tau$ (black triangles), (g) $R_d$ (blue square), $R_{ph}$ 
    (green circle), $R_s$ (red hexagon) and $R_0$ (grey diamond), (h) $N_e$ (red triangle) and $f_{N,pl}$ (grey hexagon), and (i) $F_{tot,ob}$ (red solid line) and $F_{comp,co}/F_{seed,co}$ (black squares).}
    \label{131014A_param}
\end{figure*}

\begin{figure*}[!ht]
    \centering
\includegraphics[width=\linewidth]{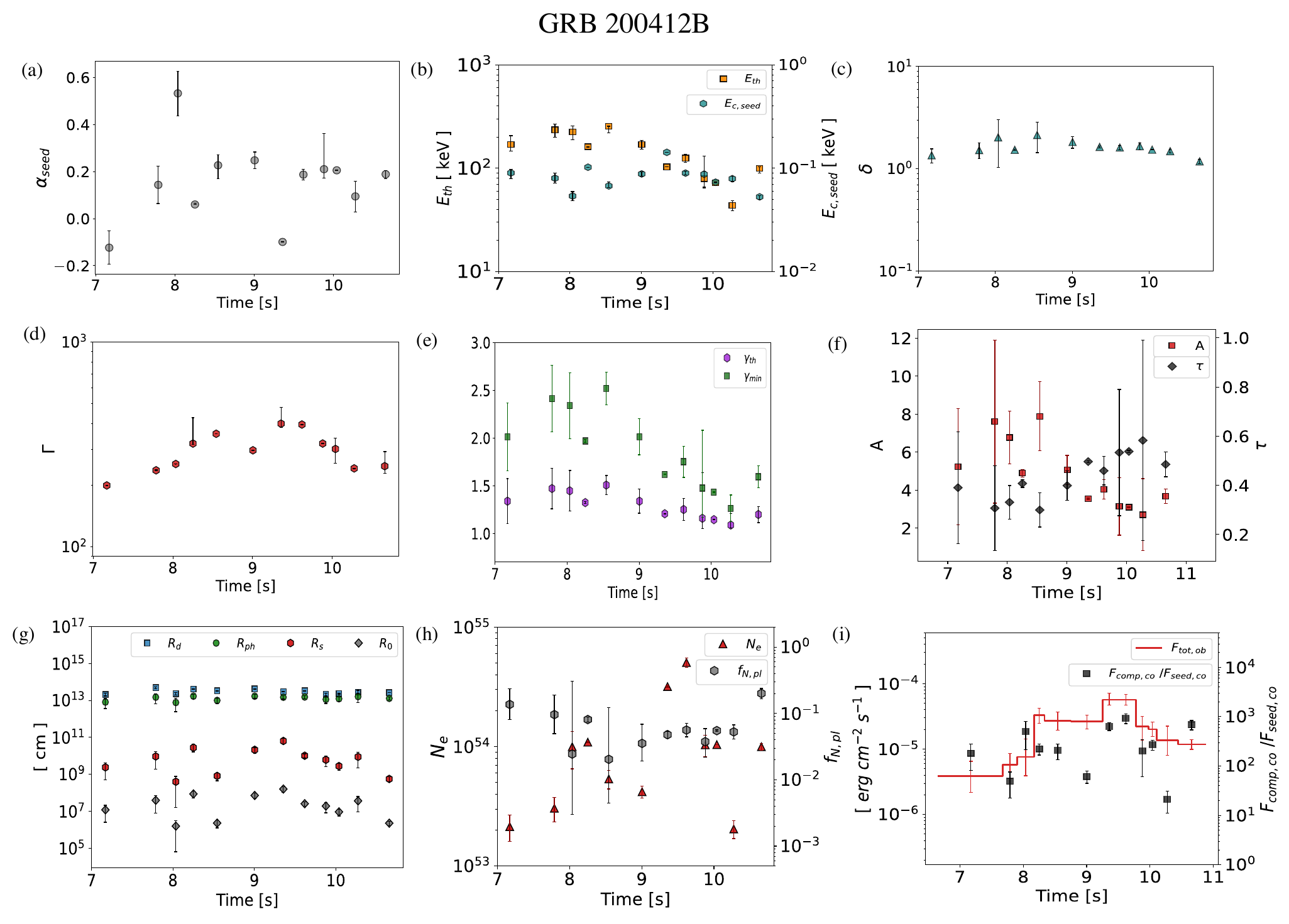}        
    \caption{The temporal evolution of the best-fit parameters of the IC model and physical parameters derived for GRB 200412B is presented in a format analogous to that of Figure \ref{131014A_param}.}
    \label{200412B_param}
\end{figure*}

\begin{figure*}[!ht]
    \centering
    \includegraphics[width=1.0\linewidth]{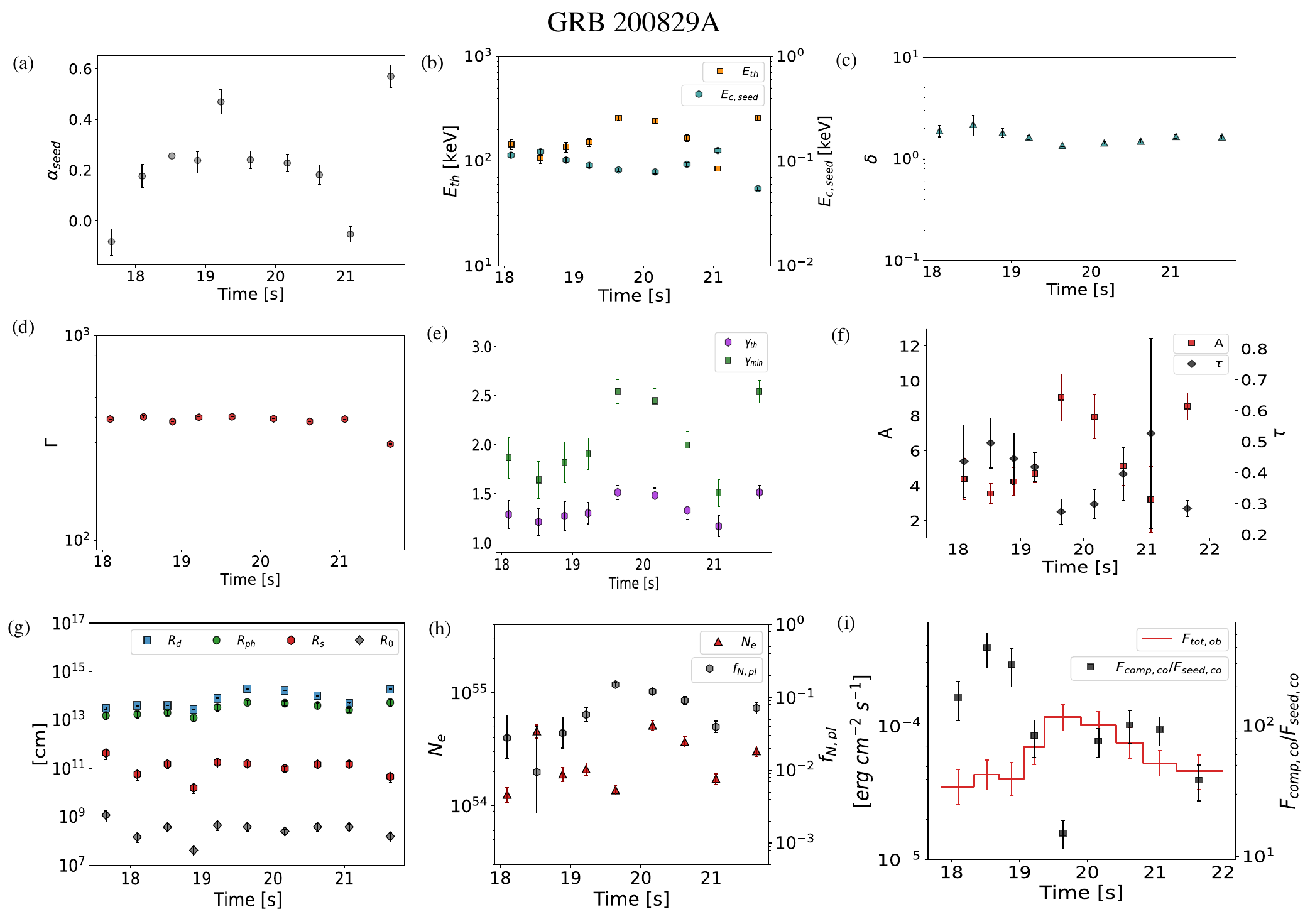}        
    \caption{The temporal evolution of the best-fit parameters of the IC model and physical parameters derived for GRB 200829A is presented in a format analogous to that of Figure \ref{131014A_param}. }
    \label{200829A_param}
\end{figure*}

\begin{figure*}[!ht]
    \centering
    \includegraphics[width=1.0\linewidth]{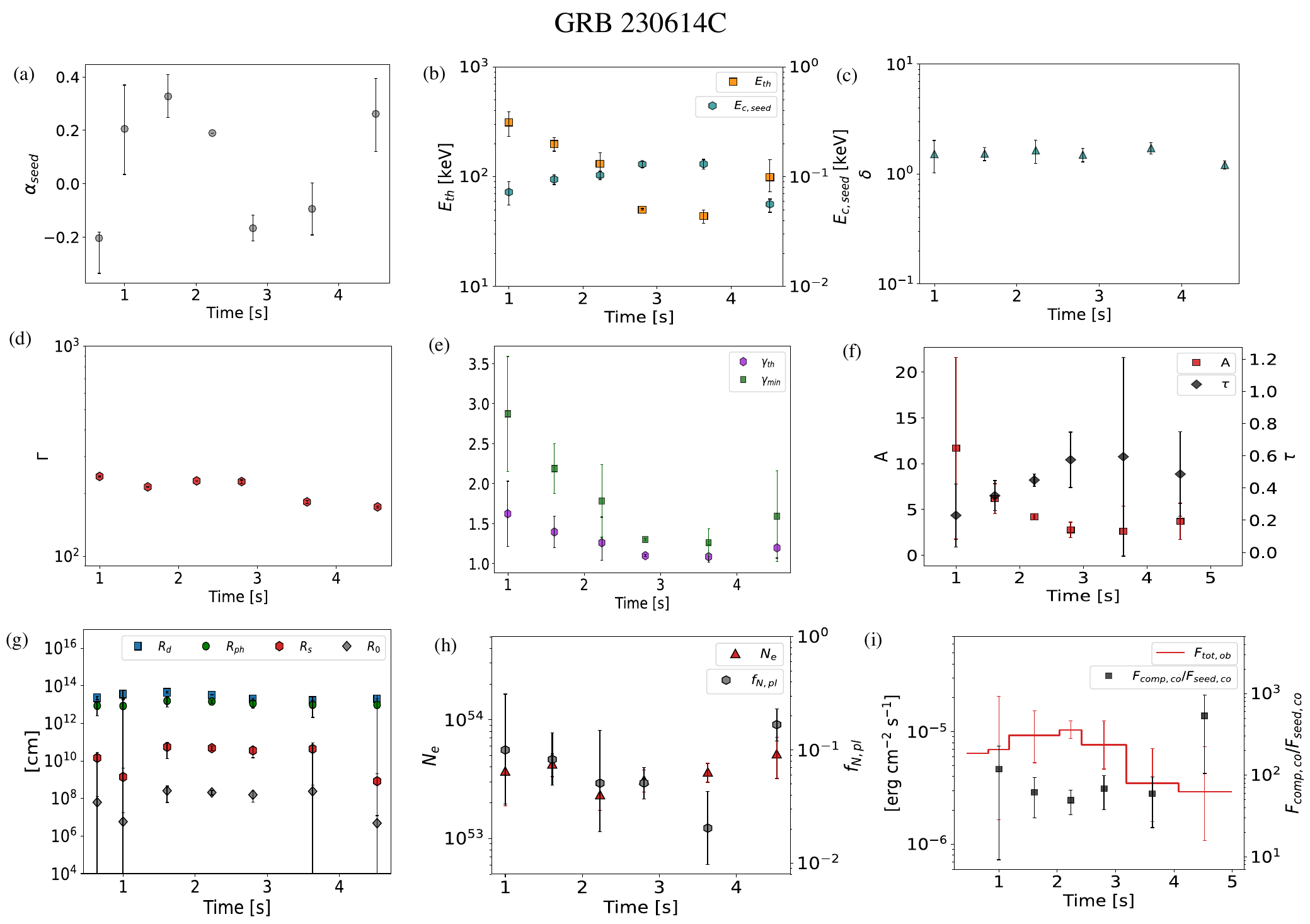}        
    \caption{The temporal evolution of the best-fit parameters of the IC model and physical parameters derived for GRB 230614C is presented in a format analogous to that of Figure \ref{131014A_param}.  }
    \label{230614C_param}
\end{figure*}

The temporal evolution of the fit parameters: $\alpha_{seed}$, $E_{c,\mathrm{seed}}$, 
$E_{\rm th}$, $\delta$, $\Gamma$, and the total observed flux $F_{\rm tot,ob}$, is presented in subpanels (a), (b, right axis), (b, left axis), (c), and (d) of 
Figure \ref{131014A_param} for GRB 131014A. 
The corresponding parameter evolutions for GRB 200412B, GRB 200829A, and GRB 230614C are shown in Figures \ref{200412B_param}, \ref{200829A_param}, 
and \ref{230614C_param}, respectively. 

Across the four GRBs in our sample, the fitted seed photon parameters show a consistent pattern. The low-energy power law index of the seed, $\alpha_{seed}$, distribution spans $-0.2$ to $+0.6$, 
indicative of a substantial contribution from unscattered thermal photons advected from the photosphere and still visible within the {\it Fermi} energy range. The mean value of the $\alpha_{seed}$ across the sample is $\approx +0.2$. The co-moving seed cutoff energy, $E_{c,\rm seed}$, 
lies between $\sim 0.05$ and $0.1$ keV, with an average value of $\approx 0.09$ keV. 

The electron distribution parameters also show broad consistency among the four bursts. 
The thermal electron energy $E_{\rm th}$ ranges from $\sim 20$ to $300$ keV with an average of $\approx 140$ keV, while the non-thermal electron power-law index $\delta$ remains relatively stable, 
typically within $1$-$3$, with an average value of $\delta \approx 1.8$.  

The inferred bulk Lorentz factors are of the order of a few hundred, spanning 
$\sim 170$ to $550$, with an average value of $\Gamma \approx 334$ across the four GRBs. 

\subsection{Inverse Compton Characteristics and Jet Outflow Parameters}
\label{derived_values}
To interpret the fitted inverse-Compton (IC) spectra in physical terms, we extract a set of quantities 
that describe the strength of Comptonisation, the location of the dissipation region, and the properties 
of the underlying electron population. In the single–scattering regime relevant for our optically thin 
IC scenario, the thermal pool of electrons produces a characteristic spectral peak, while the power-law 
tail of the electron distribution generates the high-energy non-thermal component observed in the 
spectrum.

From the fits we obtain the thermal electron energy, $E_{\rm th}$, from which the mean Lorentz factor 
of the thermal electrons follows as \citep{Bordoloi_Iyyani_2025}
\begin{equation}
    \langle\gamma_{{\rm th}}\rangle = 1 + \frac{E_{\rm th}}{m_e c^2}.
\end{equation}
The corresponding mean velocity factor is
\begin{equation}
    \langle\beta_{{\rm th}}^2\rangle = 1 - \frac{1}{\langle\gamma_{{\rm th}}^2\rangle}.
\end{equation}
\noindent
Given $\kappa =3$, the $\gamma_{min}$ is also estimated.
The temporal evolution of the estimated values of $\gamma_{\rm th}$ and $\gamma_{\rm min}$ for GRB~131014A, GRB~200412B, GRB~200829A, and GRB~230614C are shown in Figures~\ref{131014A_param}(e), \ref{200412B_param}(e), 
\ref{200829A_param}(e), and \ref{230614C_param}(e), respectively. The inferred $\gamma_{\rm th}$ values span the range $\sim 1.08-1.5$, while $\gamma_{\rm min}$ varies between $\sim 1.2-2.6$. 

The characteristic up-scattered photon energy is then
\begin{equation}
   E_{comp} = 
   \left(\frac{4}{3}\langle\gamma_{e,{\rm th}}^{2}\rangle 
   \langle\beta_{e,{\rm th}}^{2}\rangle\right) E_{c,{\rm seed}},
   \label{eq:avg_energy}
\end{equation}
which yields the mean Compton amplification factor,
\begin{equation}
    A = \frac{E_{comp}}{E_{c,{\rm seed}}},
    \label{eq:A}
\end{equation}
and the corresponding Compton $y$ parameter,
\begin{equation}
    y = \ln(A).
    \label{eq:y}
\end{equation}
\noindent
In the optically thin regime, the optical depth at the dissipation site is estimated as
\begin{equation}
    \tau = \frac{y}{A - 1},
    \label{eq:tau}
\end{equation}
from which the dissipation radius follows:
\begin{equation}
    R_{d} = \frac{R_{\rm ph}}{\tau},
    \label{eq:Rd}
\end{equation}
where $R_{\rm ph}$ is the photospheric radius inferred from the burst luminosity and the bulk Lorentz 
factor $\Gamma$.
The temporal evolution of the Comptonisation characteristics, the amplification factor $A$, the optical depth $\tau$, and the dissipation radius $R_{d}$,
for GRB~131014A, GRB~200412B, GRB~200829A, and GRB~230614C are shown in panels (f) and (g) of Figures~\ref{131014A_param}, \ref{200412B_param}, \ref{200829A_param}, and \ref{230614C_param}, respectively. Across the four bursts, the amplification 
factor spans $A \sim 2.5$--$12$, while the optical depth remains within $\tau \sim 0.2$--$0.6$. These values correspond to dissipation radii in the range $R_{d} \sim 8 \times 10^{12}\,\mathrm{cm}$ to $\sim 
10^{14}\,\mathrm{cm}$. Notably, $R_{d}$ consistently lies just above the photospheric radius in each burst, indicating that dissipation occurs in a region where the optical depth has dropped slightly below unity.

Once knowing the characteristics of the electron distribution allows us to estimate the total number of electrons, $N_e$, 
and the number of electrons in the power law tail of the electron distribution using equation 16 in \citealt{Bordoloi_Iyyani_2025}). This allows us to estimate the fraction of electrons accelerated into the power law, 
$f_{N,pl}$. Across the four bursts, the total number of electrons, $N_{e}$, is found to lie in the range 
$10^{53}$--$10^{55}$. Within each burst, the temporal evolution of $N_{e}$ typically spans no more than an order of magnitude. The fraction of electrons accelerated into the power-law tail is 
inferred to range between $f_{N,{\rm pl}} \sim 0.1\%$--$20\%$. On average across the sample, this fraction is found to be 
approximately $6\%$, which is of the same order of magnitude as the $\sim10\%$ non-thermal fraction reported in particle-in-cell simulations \citep{Spitkovsky2008}. The temporal 
evolution of $N_e$ and $f_{N,{\rm pl}}$ for the bursts GRB 131014A, GRB 200412B, GRB 200829A and GRB 230614C is shown in the panel (h) of Figure \ref{131014A_param}, \ref{200412B_param}, \ref{200829A_param} and \ref{230614C_param} respectively. 

The total observed flux $F_{tot,ob}$ is computed over the energy range 8 keV–2 GeV for GRB 131014A and over 8 keV–30 MeV for the remaining GRBs. By Doppler deboosting, the corresponding co-moving total Comptonised flux is obtained (i.e $F_{\rm tot,co} = {\cal{D}}^{-4}\, F_{tot,ob}$) which corresponds to the co-moving integrated luminosity ($L'_{\rm IC}$) as follows: 
\begin{equation}
    F_{\rm tot,co} = \frac{L'_{\rm IC}}{4\pi R_d^2}
\end{equation}
Because the fitted spectrum contains both the unscattered and Comptonised thermal components, we 
evaluate the co-moving seed flux,
\begin{equation}
F_{\rm seed,co} 
= \int_{E_{\min}'}^{E_{\max}'} 
K \left( \frac{E'}{E_0} \right)^{1+ \alpha_{seed}}
\exp\!\left( -\frac{E'}{E_{c,{\rm seed}}} \right)
\, {\rm d}E' ,
\end{equation}
in the energy range $0.01$ keV to $1$ keV; and obtain the co-moving Comptonised flux as 
$F_{\rm comp,co} = F_{\rm tot,co} - F_{\rm seed,co}$.  
Across the sample, the ratio $F_{\rm comp,co}/F_{\rm seed,co}$ typically ranges from a few tens to 
$\sim 1000$. The temporal evolution of the total observed flux, $F_{tot,ob}$,
and the ratio of the co-moving fluxes $F_{\rm comp,co}/F_{\rm seed,co}$ for the bursts GRB 131014A, GRB 200412B, GRB 200829A and GRB 230614C are shown in the panel (i) of Figure \ref{131014A_param}, \ref{200412B_param}, \ref{200829A_param} and \ref{230614C_param} respectively. 
\noindent
\\

Combining $F_{\rm tot,ob}$, $\Gamma$ and $T'(R_{ph})$ (equation \ref{comoving_Trph}) 
allows estimation of the outflow radii, including the photospheric radius $R_{\rm ph}$, the nozzle radius $R_0$, 
and the saturation radius $R_s$ following the equations in \citealt{Pe'er2007,Iyyani_etal_2015}. The temporal evolution of these different radii for the bursts GRB 131014A, GRB 200412B, GRB 200829A and GRB 230614C is shown in the panel (g) of Figure \ref{131014A_param}, \ref{200412B_param}, \ref{200829A_param} and \ref{230614C_param} respectively. 

Together, these derived quantities provide a self-consistent physical characterization of the Comptonising region and the dynamics of the GRB outflow. We further note that studies of single-pulse GRBs indicate that $\Gamma$ typically exhibits a monotonic decline, while $R_0$ increases with time to values of $\sim 10^{8}$–$10^{9}$ 
cm before stabilising or decreasing at later stages \citep{Iyyani2013,Iyyani2016}. Consistent with this behaviour, GRB 230614C, characterised by a broad single pulse, shows a mild 
monotonic decrease in $\Gamma$ from $\sim 250$ to $\sim 180$, accompanied by an increase in $R_0$ from $\sim 10^{7}$ cm to $\sim 10^{8}$ cm followed by a plateau. Overall, these trends are consistent with expectations from single-pulse studies. By 
contrast, the multi-pulsed bursts in our sample (GRB 131014A, GRB 200412B, and GRB 200829A) represent superpositions of multiple emission episodes, and therefore the derived parameters should be interpreted as average representations rather than direct analogues of single-pulse evolution.

\section{Discussion} \label{discussion}

\subsection{Parameter Degeneracies and Fitting Uncertainties}
\label{degeneracies}
In a direct physical model such as optically thin inverse Compton scattering, parameter degeneracies are unavoidable 
because several physical quantities shape the observed flux and spectrum in similar ways. Here we summarise the dominant 
degeneracies relevant to our analysis and the steps taken to mitigate them.

A primary degeneracy arises between the seed-photon normalisation $K$ and the 
electron normalisation $N_0$. Since the IC flux in the co-moving frame scales as
\begin{equation}
F'_{\rm IC} \propto N_e\, u_{\rm seed},
\end{equation}
an increase in $N_0$ can be compensated by a corresponding decrease in $K$, and 
vice versa, yielding nearly indistinguishable IC spectra.

A second degeneracy exists between the electron thermal energy $E_{\rm th}$ (which determines the thermal Lorentz factor $\gamma_{\rm th}$) and the seed-photon cutoff energy $E_{c,{\rm 
seed}}$, because the IC peak energy depends on both quantities through Eq.~\ref{eq:avg_energy}. Consequently, a higher $E_{\rm 
th}$ paired with a lower $E_{c,{\rm seed}}$ can reproduce the same peak 
location as a lower $E_{\rm th}$ and higher $E_{c,{\rm seed}}$.

A further important degeneracy involves the bulk Lorentz factor $\Gamma$. Because {\it Naima} computes the IC spectrum in the co-moving frame and the observer-frame spectrum is obtained solely 
by Lorentz boosting, variations in $\Gamma$ modify the observed spectrum through Doppler shifting and boosting. These effects can be partially compensated by adjusting $N_0$ and $K$, since 
boosting the co-moving spectrum with a larger $\Gamma$ can mimic a model with a smaller $\Gamma$ but intrinsically higher seed or electron normalisations. Thus, different combinations of 
$\Gamma$, $N_0$, and $K$ can yield similar observed spectra. This reflects the fact that the data constrain the shape and 
normalisation of the up-scattered spectrum more strongly than the unique partition between relativistic boosting and intrinsic electron and seed energies.
 
To resolve these degeneracies, we impose physically motivated bounds on $\Gamma$, $E_{c,{\rm seed}}$, and $N_0$ derived from the fireball model and constrained by the observed isotropic luminosity and plausible nozzle radii (Section~\ref{param_space}). These restrictions ensure that the fitted values of $\Gamma$ remain self-consistent with the expected dynamics of the GRB outflow.

Finally, the high-energy spectral slope of the IC component follows the asymptotic relation $N(E)\propto E^{-(\delta + 1)}$, 
meaning that constraints on the electron power-law index $\delta$ depend sensitively on the limited high-energy 
coverage of GBM above $\sim 30$~MeV (particularly when LAT data are unavailable). We therefore adopt physically motivated bounds on $\delta$, informed by theoretical expectations and previously reported values, ensuring stable and meaningful uncertainties on this parameter.

The statistical manifestation of these intrinsic degeneracies, as quantified through the covariance-derived parameter correlations across time-resolved intervals for the different GRBs, is shown in Figure \ref{correlation_analysis_plot} and discussed in Appendix \ref{fit_correlations}.

\subsection{Synchrotron vs. IC Electron Distributions - Comparison}
\label{sync_ic_comp}
The minimum electron Lorentz factor, $\gamma_{\min}$, is a key microphysical parameter characterising 
particle energisation at the dissipation site. Because it cannot be measured directly, its value must be 
inferred within the context of a specific emission model. In GRB studies this is most commonly done within 
the synchrotron framework, which successfully explains broadband afterglow emission. Depending on whether 
electrons cool faster or slower than the dynamical timescale, synchrotron modelling typically yields 
$\gamma_{\min}$ values ranging from a few hundred up to $10^{4}-10^{7}$.

Interpreting the Band $E_{\rm peak}$ as the synchrotron peak leads to a wide range of inferred electron 
Lorentz factors: for bulk Lorentz factors of a few hundred, very strong magnetic fields 
($B \sim 10^{6} -10^{7}$ G) imply $\gamma_e \sim 10^2$, while much weaker fields 
($B \sim 10$ - $10^{-4}$ G) yield $\gamma_e \sim 10^{4}-10^{7}$ \citep{Iyyani2016}. Furthermore, \citet{Burgess2014a} showed that most GRB prompt spectra are incompatible with fast-cooling 
synchrotron and instead favour slow-cooling emission. This requires intrinsically large $\gamma_e$ values and 
implies that either only a small fraction of electrons are accelerated to ultra-relativistic energies or the 
emission arises from forward shocks \citep{Panaitescu_Meszaros1998,Burgess_etal_2016}. In addition, slow cooling typically implies low radiative efficiencies. Although heating–cooling balance can in principle produce a steady-state distribution, such conditions are generally 
not expected in baryonic internal shocks.

A recent investigation by \citealt{Oganesyan_etal_2019}, directly fitted synchrotron spectra to prompt optical + X–ray + $\gamma$ ray data and found that achieving acceptable fits within a standard synchrotron prompt-
emission scenario requires unusually low magnetic fields, very high $\gamma_m$ values, and dissipation radii exceeding $10^{16}$ cm (the typical deceleration radius), conditions 
that favour marginally fast cooling but are difficult to reconcile with conventional internal dissipation models. More recently, \cite{Yan_2024} investigated the applicability of synchrotron model to a sample of single-
pulsed GRBs. In their framework, the emission arises from a single shell launched by the central engine, characterised by fixed outflow properties such as the bulk Lorentz factor ($\Gamma$) and nozzle radius ($R_0$), along with a single 
injection of electrons into a power-law distribution that cool over time in a decaying magnetic field as the emission shell expands. By inferring an initial emission region of $\sim 10^{15}$ cm, most of the bursts indicate low inferred magnetic 
field strengths (of order $\sim 10$ G and lesser) and high minimum electron Lorentz factors ($\gamma_{min} \sim 10^{4}–10^{6}$) for most of the burst duration. The bulk Lorentz factors are found to be of the order of a few hundred, broadly consistent with our findings.

In contrast, the optically thin inverse Compton (IC) scenario explored in this work allows the initial physical 
conditions at the central engine to evolve dynamically with time, without imposing any specific functional form on their temporal behaviour. In this framework, we consider a hybrid electron 
distribution, consisting of a Maxwellian component combined with a power-law tail. The physical properties are inferred through time-resolved spectral fitting, treating each time interval as independent. Within this framework, we find that the minimum electron Lorentz factor, $\gamma_{\min}$, can take far more modest values at an emission (dissipation) radius of 
$\sim 10^{14}$ cm. Such a hybrid electron distribution considered in this work is supported by PIC simulations (e.g.\ \citealt{Spitkovsky2008}). The thermal electron pool is characterised by $\gamma_{th} \approx 1.3$, indicating that the electrons reaching the dissipation site are 
only mildly relativistic and remain close to a cold, sub-relativistic regime. Only a small fraction ($\sim 10\%$) of electrons are accelerated into a non-thermal tail 
with index $\delta \approx 1.8$, and the minimum Lorentz factor of this component is only $\gamma_{\min} \sim 2$. However, the non-thermal tail can extend to $\gamma_{\max} \sim 10^{4}$, as shown by \citet{Bordoloi_Iyyani_2025}. 

These results point toward dissipation mechanisms that naturally produce heated Maxwellian distributions of 
mildly relativistic electrons. A particularly relevant example is the slow-heating scenario 
\citep{Ghisellini&Celotti1999, Pe'er2006}, in which, just above the photosphere ($\tau_{\gamma e} \lesssim 1$), 
the jet's kinetic energy is gradually dissipated through plasma instabilities and turbulence. This allows 
electron energisation over timescales comparable to the dynamical time, in contrast to the impulsive 
acceleration characteristic of internal shocks \citep{Kobayashi_etal_1997, Sari1998}. Turbulence-driven slow 
heating naturally yields Maxwellian electrons with $\gamma_{e,{\rm th}} \sim 2$--5, with a small fraction 
accelerated into a power-law tail beginning at similarly low $\gamma_{\min}$.

Thus, these contrasting values of $\gamma_{\min}$ highlight that the inferred electron energies are strongly radiation model dependent. Synchrotron interpretations typically require highly 
relativistic electrons, consistent with strong shock acceleration, whereas the IC interpretation presented here 
points to mildly relativistic, slowly heated electrons just above the photosphere. Hence, the characteristic electron 
Lorentz factor becomes a key discriminator between competing emission models and the physical conditions at the dissipation site.

\subsection{Physical Implications: Spectral Fit $\&$ Derived Parameters}
Based on the derived IC parameters, we find that the Compton amplification factor spans values from a few up to a ten, corresponding to 
$y = \ln(A) \sim 1$--$3$. These moderate $y$ values place the emission within the unsaturated inverse-Compton regime. The 
electron power-law index remains relatively stable across the sample, with an average of $\delta \simeq 1.8$, consistent with expectations from first-order \textit{Fermi} acceleration. 
The dissipation is inferred to occur just above the photosphere, at optical depths of $\tau \sim 0.2$--$0.6$. The outflow and 
dissipation parameters obtained here closely resemble those empirically inferred in \citet{Bordoloi_Iyyani_2025}. 

In such mildly optically thin environments, the compactness parameter is expected to satisfy $l > 1$, allowing photon--photon annihilation to generate 
electron--positron pairs. This increases the effective particle number and the opacity, while still retaining overall optical thinness \citep{Bordoloi_Iyyani_2025}. The enhanced particle 
population leads to a reduction in the average electron energy and naturally suppresses higher-order scatterings. This behaviour supports the physical validity of employing a single-
scattering IC model such as the one implemented in \textit{Naima}. A comprehensive discussion of this regime is provided in Section 5.4 of \citet{Bordoloi_Iyyani_2025}. 

Moreover, the observed GRB spectra exhibit a curved $\nu F_{\nu}$ peak accompanied by a distinct high-energy break 
and an extended power-law tail. While the majority of prompt GRB spectra are adequately described by single-break functions such as the Band function 
\citep{Band1993}, power law with exponential cutoff, or smoothly broken power law \citep{GBMcatalog2014}, the additional high-energy structure seen in these events requires a more complex empirical model combination.

Such more complex shapes at higher energies become more evident when LAT LLE and LAT $>100$~MeV data provide stronger constraints. In a few bright GRBs, complex spectral 
behaviour has been reported, including an extra power-law component extending to higher energies \citep{Abdo2009_090902B,Ryde2010,Ryde2011,Iyyani_etal_2015,Sharma_etal_2019}, a power law with an exponential cutoff \citep{Ackermann2011}, or the presence of multiple breaks 
demonstrated using models such as two smoothly broken power laws \citep{Ravasio_etal_2018,Ravasio_etal_2019A,Toffano_etal_2021}. GRB~131014A has also been shown to exhibit similar spectral complexity \citep{Bordoloi_Iyyani_2025}. 
Some cases even show multiplicative components, such as an exponential cutoff at higher energies \citep{Ackermann2011, Vianello2018, Sharma_etal_2019}.  

The acceptable IC physical fits obtained for the GRBs in our sample provide a natural interpretation for this behaviour, as 
inverse Compton scattered spectrum readily produces curved peaks closely tracks the underlying electron distribution, such that a 
predominantly mildly heated electron population with a small non-thermal tail naturally reproduces the observed curved peak, high-energy break, and extended power-law component.
The successful IC fits therefore point toward a photon-dominated, low-magnetic-field dissipation region in which the electrons are predominantly mildly heated, with only a small fraction undergoing strong non-thermal acceleration.

For comparison, most direct physical synchrotron models developed so far have focused on reproducing the general single-peaked prompt GRB spectra 
\citep{Burgess2014a,Zhang_etal_2016,burgess_etal_2019NatAs,Oganesyan_etal_2019}, and only a few attempts have been made to account for more complex shapes like broader spectra with high-energy cutoffs, e.g in case of 
GRB~160821A \citep{Ryde_etal_2022}. These efforts demonstrate that synchrotron-based models can be extended toward more complex regimes, which may require additional assumptions 
regarding magnetic-field evolution, cooling, or electron acceleration efficiency. 


The inference of mildly relativistic electron Lorentz factors, $\gamma_{\min}$ within the IC model, 
indicates that the energetic demands placed on the electron population are significantly lower. Consequently, an IC-dominated radiative scenario suggests that the mechanism responsible for dissipating the jet’s kinetic energy may operate with only modest efficiency, making it an energetically efficient channel for producing the prompt emission in these bursts.

\subsection{Thermal Emission Constraints from IC Modelling}
\label{thermal_constraint}
In selecting our sample, we specifically required $\alpha > -0.45$, ensuring that the low-energy part of the observed 
{\it Fermi} GBM spectrum is dominated by the photospheric thermal emission. This criterion guarantees that the 
unscattered thermal component remains within the GBM bandpass and is therefore directly observable. Indeed, for several 
bursts in our sample—GRB~131014A, GRB~200412B, and GRB~200829A—the empirical fits showed a statistically significant 
blackbody (BB) component when comparing the Band and Band+BB models. GRB~230614C, while satisfying $\alpha > -0.45$, did not 
show strong evidence for an additional BB component, though its spectral shape still indicates the presence of a thermal 
contribution.

In the regime where the thermal (photospheric) component lies within the {\it Fermi}–GBM energy window, the fitted seed-photon distribution obtained from the {\it Naima} IC model must be 
interpreted as representing the total thermal photon field present at the dissipation site, i.e. the sum of unscattered and 
scattered photons. This arises from the way the {\it Naima} IC formalism performs the convolution: the entire input seed 
spectrum is assumed to interact with the electron distribution, and every photon in the seed field is treated as available for scattering.
Because we do not include an independent blackbody component in addition to the IC model when fitting the data, {\it Naima} must reproduce both (i) the observed low-energy thermal peak and (ii) the high-energy Comptonised (non-thermal part of the spectrum) using a single seed field. In 
practice, this means that part of the seed distribution is effectively “left unscattered” by the model - physically corresponding to photons undergoing only Thomson-like 
scatterings in the co-moving frame, which do not significantly alter their energy. The same seed field is then also upscattered by the relatively energetic part of the electron distribution to reproduce 
the high-energy IC emission modelling the non-thermal part of the spectrum.
Thus, the seed photon normalisation and temperature inferred from the Naima IC fit encode both the unscattered photospheric flux and the fraction of seed photons that have undergone energy-
boosting via inverse Compton scattering. This interpretation is appropriate only in cases where the thermal component is 
directly visible in the data, as in the GRBs analysed in this work.

In contrast, when the photospheric emission thermal temperature lies below the GBM energy range, the observed spectrum may consist only of the Comptonised component. In such situations, 
the empirical low-energy slope typically appears softer (closer to $\alpha \sim -1$).  A similar scenario involving IC in sub-photospheric dissipation scenarios had been presented in 
\citealt{Ahlgren_etal_2019}. These kind of spectra were excluded from our study because they can be degenerate with synchrotron interpretations. Nevertheless, the IC model can still physically 
account for such cases: the thermal peak would then lie below the instrumental threshold, and the {\it Naima} IC fit would yield the scattered thermal component responsible for the 
observed IC spectrum. This allows recovery of the co-moving 
thermal flux that produced the Comptonised emission, although the unscattered thermal flux remains unconstrained in these situations.

\section{Summary and Conclusions} 
\label{conclusion}
In this work, we have carried out a systematic investigation of the prompt emission of four GRBs using a direct, physically motivated model of optically thin inverse Compton (IC) 
scattering implemented through the {\it Naima} framework. Starting from a catalog-based selection of 41 bright {\it Fermi} GRBs, we identified four events whose time-resolved spectra 
satisfy physically motivated criteria [$\alpha > -0.45$ and $-1.7 >\beta > -3.3$] that make them suitable for direct IC modelling. Using a customised {\it Naima} based IC framework implemented within 
the 3ML fitting environment, we carried out fully forward-folded spectral fits to the GBM (and where applicable, LLE/LAT) data. For each GRB, we extracted the temporal evolution of the key model parameters, seed-photon characteristics, electron 
distribution properties, and the bulk Lorentz factor, and used these to derive physically relevant quantities such as the Compton amplification factor, optical depth, dissipation radius, 
and electron population statistics. The combination of well constrained spectral fits, goodness-of-fit testing in terms of obtained random fit residuals, and physical consistency checks demonstrates that an IC-dominated emission scenario can 
successfully reproduce the observed prompt spectra in these bursts.

The derived parameters reveal a coherent physical picture across the sample. The Compton $y$-parameter lies firmly in the 
unsaturated regime, with amplification factors of tens to thousands, while the inferred dissipation radii reside just above the photosphere at optical depths $\tau \sim 0.2{-}0.6$. 
The electron population is characterised by mildly relativistic thermal Lorentz factors and a non-thermal tail with a typical index, $\delta \approx 1.8$, implying an energetically modest and photon-dominated dissipation environment. The IC model naturally accounts for the observed spectral curvature and high-
energy extension without requiring fine-tuned magnetic fields or extreme electron energies, distinguishing it from synchrotron only interpretations. Moreover, the framework enables constraints on sub-dominant thermal emission even when it lies below the detector’s energy threshold. Overall, this study demonstrates that optically thin IC scattering 
provides a physically self-consistent and observationally successful description of the prompt emission in a subset of bright GRBs, and motivates broader application of physically grounded IC models in future GRB analyses.

\begin{acknowledgments}
S.I. is supported by DST INSPIRE Faculty Scheme (IFA19-PH245) and SERB SRG Grant (SRG/2022/000211). This
research has made use of {\it Fermi} data obtained through High Energy Astrophysics Science Archive Research Center Online Service, provided by the NASA/Goddard Space Flight Center. This work utilized various software such as 3ML, NAIMA, PYTHON (\citealt{Python_1}), ASTROPY (\citealt{Astropy_2013,Astropy_2018,Astropy2022}), NUMPY (\citealt{Numpy2011,Numpy_1}), SCIPY (\citealt{Scipy2001,Scipy_1}), MATPLOTLIB (\citealt{Matplotlib_1}), FTOOLS (\citealt{ftools_1}) etc. We acknowledge the support of High Performance Computing Centre (CHPC) of IISER TVM for providing the computational resources. 
\end{acknowledgments}

%






\appendix 

\section{Summary of the catalog of 41 GRBs}
\label{comparison}
\begin{figure*}[!ht]
    \centering
    \includegraphics[width=\linewidth]{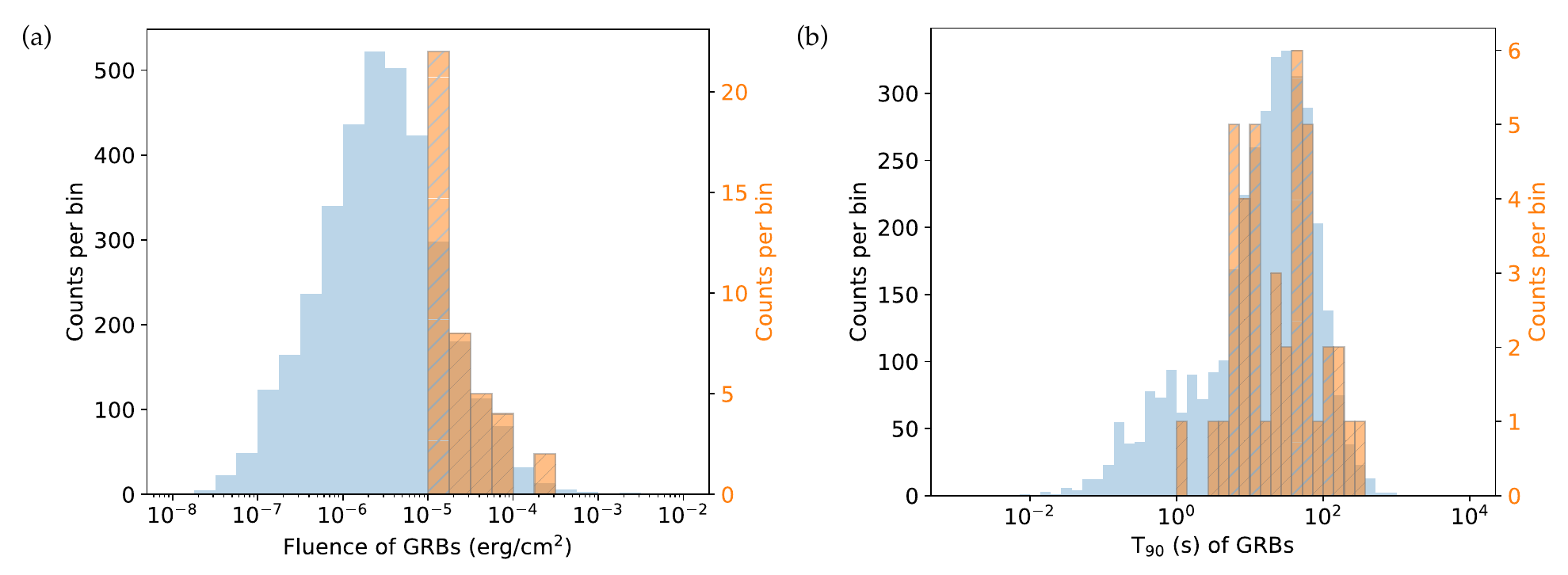}        
    \caption{The comparison of the initial sample of 41 GRBs (orange hatched) with the {\it Fermi} GBM (blue) spectral catalog in terms of (a) Fluence and (b) $T_{90}$ are shown.}
    \label{comp_fluence_t90}
\end{figure*}

\begin{figure*}[!ht]
    \centering
    \includegraphics[width=\linewidth]{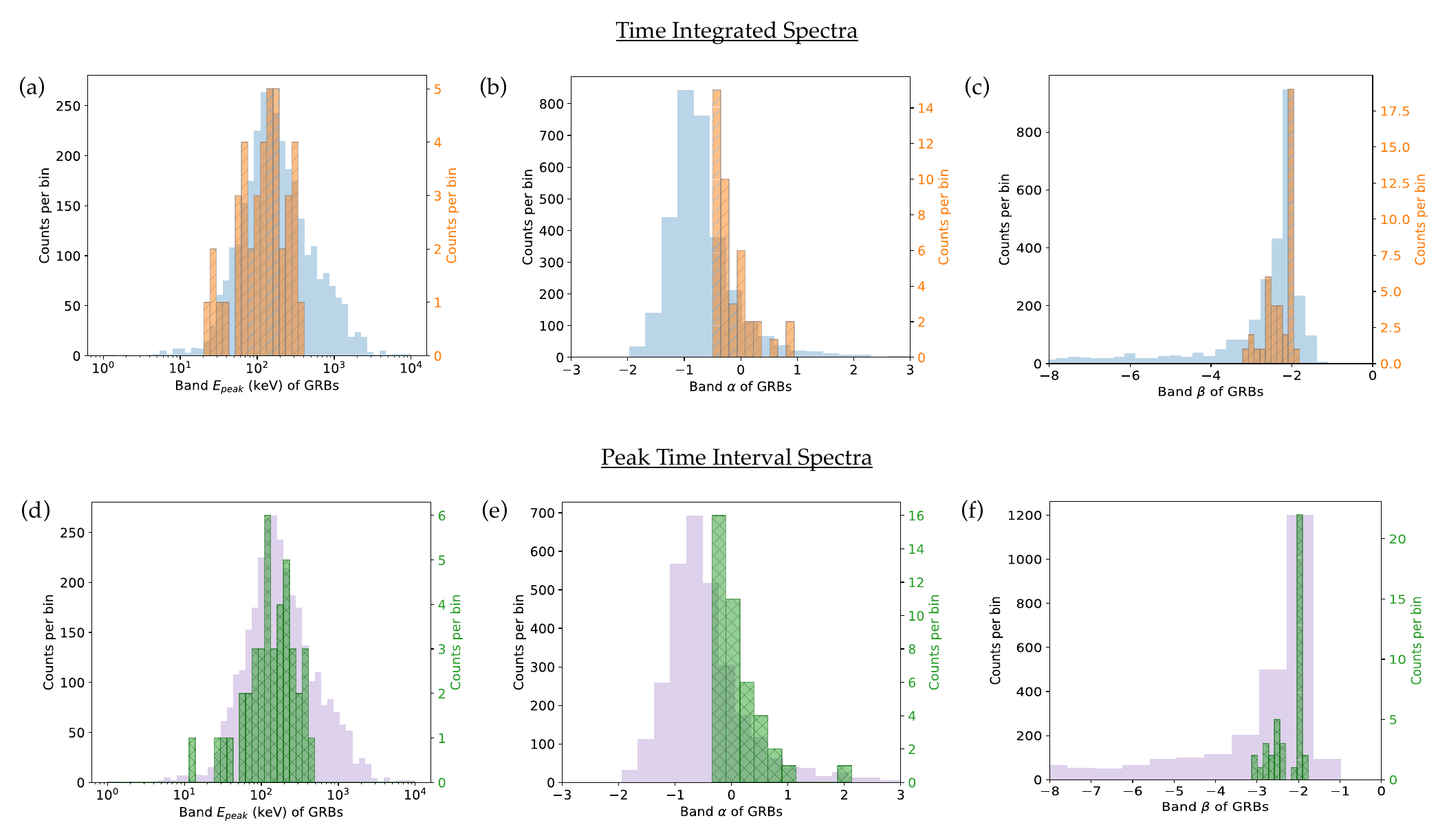}        
    \caption{The comparison of the initial sample of 41 GRBs (orange/green hatched) with the Fermi-GBM (blue/purple) spectral catalog is shown for the Band function fit parameters $E_{peak}$, $\alpha$, and $\beta$ in both the time-integrated [(a), (b), (c)] and peak-interval [(d), (e), (f)] spectra are shown.}
    \label{comp_integ_peak}
\end{figure*}

A comparison between our sample of 41 GRBs and the full {\it Fermi} GBM spectral catalog shows that the selected bursts occupy the brighter and more energetic end of the population. As seen in 
the $T_{90}$ –fluence distribution (Figure \ref{comp_fluence_t90}), the sample predominantly consists of moderately long GRBs with fluences typically above 
$10^{-5}\,\mathrm{erg\,cm^{-2}}$, reflecting the intentional selection of bright events suitable for detailed time-resolved spectroscopy. In terms of spectral properties, the distributions 
of Band $\alpha$, $\beta$ and $E_{\rm peak}$ for both time-integrated and peak-interval fits lie within the typical ranges observed in the full catalog, but are skewed toward harder low-
energy slopes and well-constrained high-energy indices (Figure \ref{comp_integ_peak}). This is because of our selection criteria requiring $\alpha > -0.5$ (time-integrated) and $\alpha 
> -0.45$ (peak), which preferentially includes bursts where a strong thermal component is visible within the GBM bandpass. Overall, the selected GRBs represent a bright, spectrally well-
characterised subsample of the GBM population, optimal for testing physical models such as the optically thin inverse-Compton scenario. 

\section{Time Resolved Spectral Fits of Band Function}
\label{sample}
\begin{figure*}
    \includegraphics[width=\linewidth]{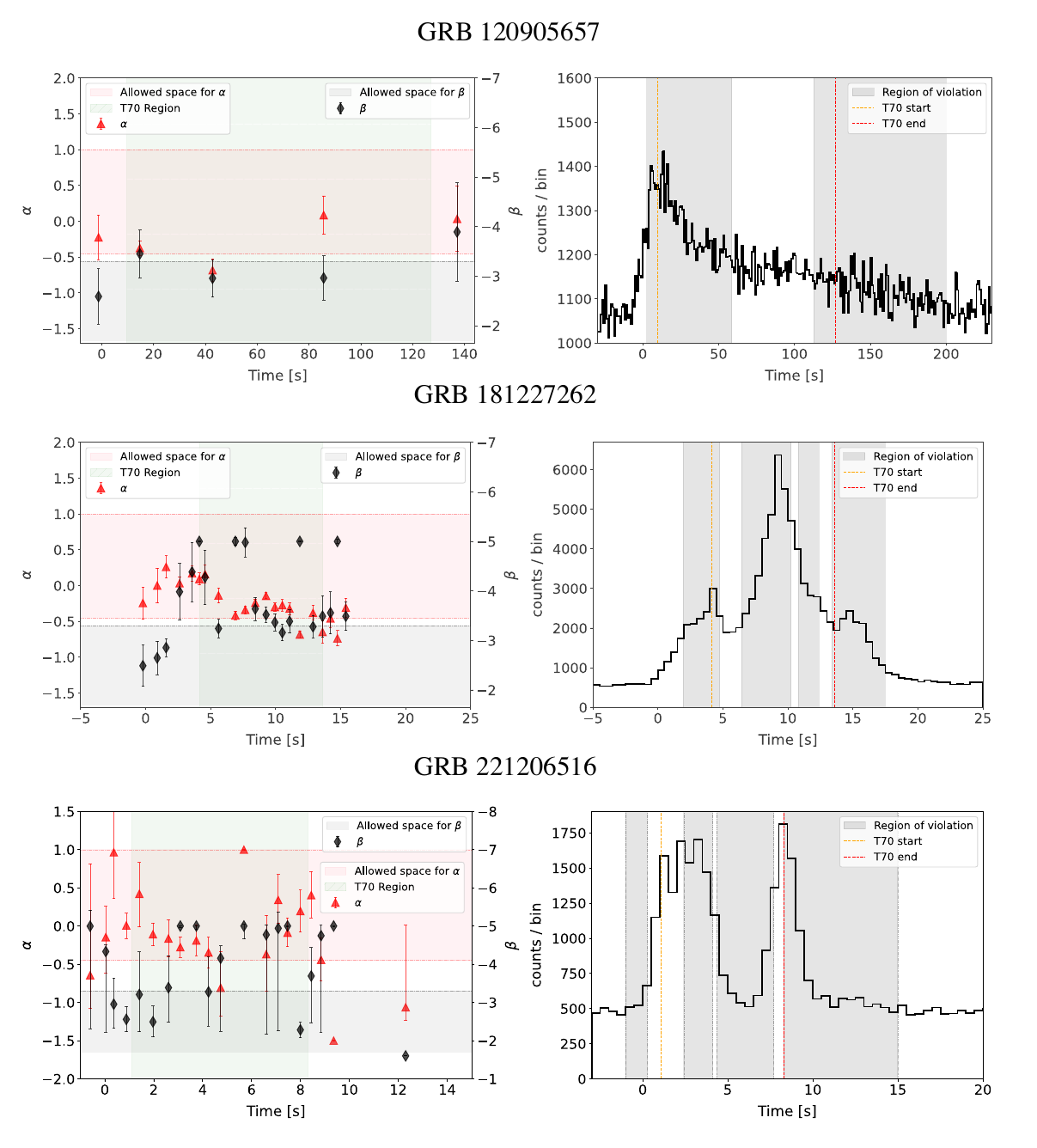}
    \caption{A few representative GRBs that were excluded from the initial sample because their time resolved Band $\alpha$ and $\beta$ did not meet the selection criteria are shown above. The complete figure set of 41 GRBs is available in the online journal.}
    \label{rejected}
\end{figure*}

As supplementary material to this work, we provide the full set of Band function spectral fit results obtained from the time-
resolved analysis of all 41 GRBs in our initial sample. The complete results are supplied as machine-readable table. A representative excerpt of the dataset including $\alpha$, $\beta$, Band peak energy, $E_p$, and normalisation ($Norm$), is given in Table \ref{tab:grb090620400_band}. We further include figure sets showing the temporal evolution of 
the Band $\alpha$ and $\beta$ parameters for each burst, highlighting the $T_{70}$ interval and the allowed ranges of $\alpha$ and $\beta$ used in the selection criteria described in 
Section~\ref{criteria}. A representative example of these figure sets is shown in Figure~\ref{rejected}.


\begin{deluxetable*}{ccccccc}
\tablecaption{A representative excerpt of the time-resolved Band function spectral fit parameters from the sample of 41 GRBs is provided below. The table illustrates the results for GRB 090620400, while the complete dataset is included in the supplementary material. The CSV file containing the Band function fit results for the entire sample is available online.} \label{tab:grb090620400_band}
\tablehead{
\colhead{Interval} &
\colhead{$T_\mathrm{start}$} &
\colhead{$T_\mathrm{stop}$} &
\colhead{$\alpha$} &
\colhead{$\beta$} &
\colhead{$E_\mathrm{p}$} &
\colhead{Norm} \\
\colhead{} &
\colhead{(s)} &
\colhead{(s)} &
\colhead{} &
\colhead{} &
\colhead{(keV)} &
\colhead{(ph cm$^{-2}$ s$^{-1}$ keV$^{-1}$)}
}
\startdata
1  &  -0.998 &  0.857 & $ 0.232^{+0.275}_{-0.283}$ & $-4.52^{+2.33}_{-0.041}$ & $238^{+27.0}_{-25.8}$  & $0.026^{+0.008}_{-0.006}$ \\
2  &   0.857 &  3.093 & $ 0.249^{+0.127}_{-0.123}$ & $-2.88^{+0.272}_{-0.258}$ & $137^{+7.36}_{-7.47}$ & $0.175^{+0.036}_{-0.029}$ \\
3  &   3.093 &  4.077 & $-0.158^{+0.100}_{-0.101}$ & $-2.61^{+0.238}_{-0.238}$ & $216^{+16.4}_{-15.0}$ & $0.102^{+0.014}_{-0.012}$ \\
4  &   4.077 &  4.796 & $ 0.020^{+0.089}_{-0.088}$ & $-3.22^{+0.508}_{-0.502}$ & $229^{+12.9}_{-12.1}$ & $0.167^{+0.019}_{-0.016}$ \\
5  &   4.796 &  5.319 & $ 0.098^{+0.130}_{-0.133}$ & $-2.85^{+0.324}_{-0.323}$ & $167^{+11.3}_{-10.5}$ & $0.202^{+0.037}_{-0.032}$ \\
6  &   5.319 &  7.452 & $-0.253^{+0.097}_{-0.096}$ & $-2.49^{+0.142}_{-0.140}$ & $128^{+7.91}_{-7.47}$ & $0.129^{+0.022}_{-0.019}$ \\
7  &   7.452 &  8.740 & $-0.413^{+0.133}_{-0.125}$ & $-3.96^{+1.50}_{-0.536}$ & $108^{+6.87}_{-7.50}$ & $0.092^{+0.023}_{-0.017}$ \\
8  &   8.740 &  9.747 & $-0.623^{+0.296}_{-0.287}$ & $-2.50^{+0.360}_{-0.380}$ & $79.1^{+14.1}_{-11.7}$ & $0.055^{+0.043}_{-0.024}$ \\
9  &   9.747 & 12.318 & $-0.413^{+0.598}_{-0.525}$ & $-2.14^{+0.189}_{-0.198}$ & $63.9^{+15.9}_{-15.3}$ & $0.060^{+0.153}_{-0.040}$ \\
10 &  12.318 & 19.000 & $-1.50^{+0.006}_{-0.001}$ & $-2.08^{+0.237}_{-0.275}$ & $75.2^{+34.7}_{-20.7}$ & $0.004^{+0.000}_{-0.000}$ \\
\enddata
\tablecomments{Uncertainties are 1$\sigma$. Times are relative to the GBM trigger, and the Band normalization
(Norm) is evaluated at 100~keV.}
\end{deluxetable*}

\section{Parameter Correlations and Degeneracy Structure in Time-resolved Spectral Fits}
\label{fit_correlations}

\begin{figure*}[!ht]
    \centering
    \includegraphics[width=\linewidth]{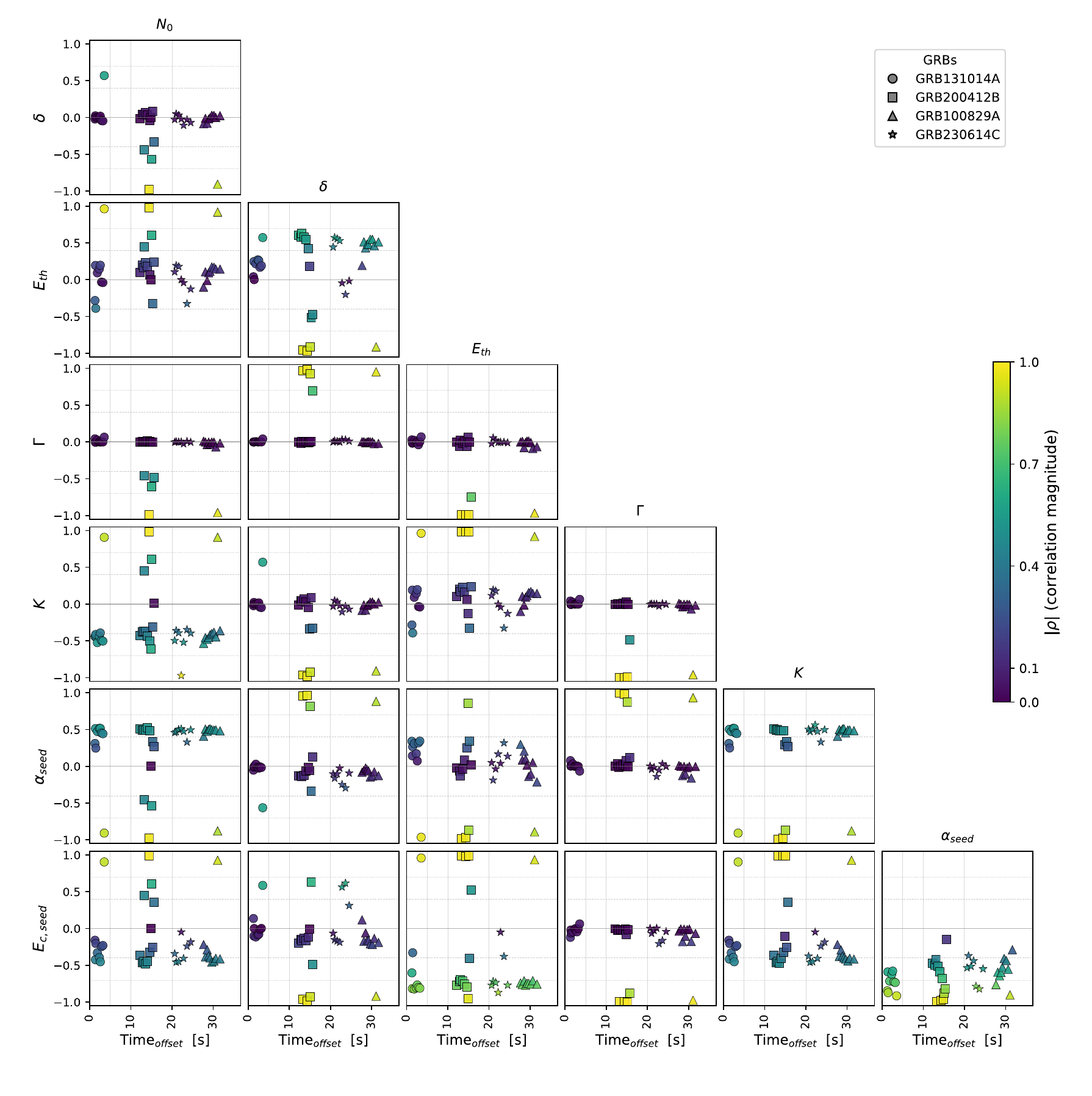}        
    \caption{The pairwise parameter correlations derived from the covariance matrices of time-resolved spectral fits of the four GRBs in our sample are shown above. Each panel shows the evolution of the correlation coefficient between a given parameter pair across spectral intervals. Marker shape identifies the GRB, while marker colour represents the magnitude of the correlation coefficient, as indicated by the colour bar. The shaded grey horizontal dashed and dotted lines mark approximate transitions between weak, moderate, and strong correlation regimes. For clarity, the time axes of individual GRBs are offset horizontally, preserving intra-burst temporal evolution while enabling direct comparison of correlation behaviour across bursts.}
    \label{correlation_analysis_plot}
\end{figure*}

Parameter degeneracy occurs when different combinations of parameters yield nearly identical likelihood values, producing extended likelihood contours in parameter space. 
Since the covariance matrix captures the local curvature of the likelihood surface around the best-fit point, and its normalised form, the Pearson correlation matrix, reveals these degeneracy directions. Consequently, large-magnitude off-diagonal correlation coefficients indicate strong 
parameter degeneracies. Figure \ref{correlation_analysis_plot} presents the temporal evolution of these pairwise correlations ($\rho$) for the four GRBs in our sample, thereby providing an empirical diagnostic of degeneracy directions and their variability 
across spectral intervals. Marker shape denotes GRB identity, while colour represents the magnitude of the correlation coefficient, 
illustrating negligible ($\rho<0.1$), weak ($\rho \sim 0.1-0.4$), moderate ($\rho \sim 0.4-0.7$), and strong ($\rho > 0.7$) degeneracy regimes.  

The correlation structure of the fitted parameters exhibits interval-dependent variability, with both the magnitude and 
sign of pairwise correlations evolving across time-resolved spectra. These correlations derived from the covariance matrix evaluated at the maximum-likelihood solution reflect the local 
likelihood geometry constrained by the data in each interval. While the inverse Compton model predicts intrinsic degeneracies between specific parameter pairs, most notably between the seed and electron normalisations and among parameters governing the spectral 
peak, their statistical manifestation depends on photon statistics, spectral curvature, and instrumental sensitivity. Intervals with well-constrained peak properties or normalisation display stronger 
correlations, whereas lower signal-to-noise time intervals show weaker signatures. Occasional sign reversals indicate changes in the local likelihood geometry rather than a breakdown of the 
underlying physical degeneracy. Persistent moderate to strong correlations therefore identify intrinsic degeneracy directions, while interval to interval variability primarily reflects evolving statistical constraints. 

Consistent with these expectations, the seed and electron normalisations ($K$–$N_0$) exhibit a predominantly moderate to strong negative correlation across bins, reflecting the IC flux scaling and representing a robust intrinsic degeneracy. 
Similarly, the peak-setting parameters ($E_{c,seed}$–$E_{\rm th}$) show strong anti-correlation, indicative of their joint control over the IC peak energy, with the correlation 
strength varying according to the quality of peak constraints. The bulk Lorentz factor and normalisations ($\Gamma$–$K$, $\Gamma$–$N_0$) largely display weak correlations. 
The seed spectral shape parameters ($\alpha_{seed}$–$E_{c,seed}$) also show a pronounced anti-correlation, corresponding to the expected shape degeneracy of the cutoff power-law representation; this effect is mitigated by imposing physically motivated 
bounds on the cutoff. All remaining parameter pairs exhibit either negligible or weak to moderate correlations, indicating limited coupling within the model.

Overall, the impact of these degeneracies is controlled through the physically motivated parameter bounds described in Section~\ref{param_space}, which restrict unphysical combinations and ensure that the inferred IC model parameters remain robust despite residual correlations.

\bibliography{Ref1}

\end{document}